\documentclass{emulateapj}
\usepackage{times,mathptm}
\usepackage{lscape}

\usepackage{epsf}
\usepackage{psfig}
\usepackage{lscape}

\def\etal{{et\,al.\,}}

\def\ltsima{$\; \buildrel < \over \sim \;$}
\def\simlt{\lower.5ex\hbox{\ltsima}}
\def\gtsima{$\; \buildrel > \over \sim \;$}
\def\simgt{\lower.5ex\hbox{\gtsima}}

\begin{document}
\title{Weighing the Black Holes in $z\approx$~2 Submillimeter-Emitting Galaxies\\ Hosting Active Galactic Nuclei}   %%% Fill in title

\author{D.M. Alexander,\altaffilmark{1} W.N.~Brandt,\altaffilmark{2}
  I.~Smail,\altaffilmark{3} A.M.~Swinbank,\altaffilmark{3}
  F.E.~Bauer,\altaffilmark{4} A.W.~Blain,\altaffilmark{5}
  S.C.~Chapman,\altaffilmark{6}\\ K.E.K.~Coppin,\altaffilmark{3}
  R.J.~Ivison,\altaffilmark{7,8} and
  K.~Men{\'e}ndez-Delmestre\altaffilmark{5}} %%% Fill in author names

\affil{$^1$Department of Physics, Durham University, Durham DH1 3LE,
  UK} %%% Fill in author affiliations

\affil{$^2$Department of Astronomy and Astrophysics, Pennsylvania
State University, 525 Davey Laboratory, University Park, PA 16802,
USA}

\affil{$^3$Institute for Computational Cosmology, Durham University, South Road, Durham DH1 3LE, UK}

\affil{$^4$Chandra Fellow, Columbia Astrophysics Laboratory, Columbia
  University, Pupin Laboratories, 550 West 120th Street, New York, NY 10027, USA}

\affil{$^5$California Institute of Technology, Pasadena, CA 91125,
USA}

\affil{$^6$Institute of Astronomy, Madingley Road, Cambridge CB3 0HA,
UK}

\affil{$^7$Astronomy Technology Centre, Royal Observatory, Blackford Hill, Edinburgh EH9 3HJ, UK}

\affil{$^8$Institute for Astronomy, University of Edinburgh, Blackford Hill, Edinburgh EH9 3HJ, UK}

\shorttitle{WEIGHING THE BLACK HOLES IN $z\approx$~2 SUBMILLIMETER EMITTING GALAXIES}

\shortauthors{ALEXANDER ET AL.}

%
%%%%%%%%%%%%%%%%%%%%%%%%%%%%%%%%%%%%%%%%%%%%%%%%%%%%%%%%%%%%%%%%%%%%%%
\begin{abstract} %%% Abstract to run on from here.
%%%%%%%%%%%%%%%%%%%%%%%%%%%%%%%%%%%%%%%%%%%%%%%%%%%%%%%%%%%%%%%%%%%%%%
%

  We place direct observational constraints on the black-hole masses
  ($M_{\rm BH}$) of the cosmologically important $z\approx$~2
  submillimeter-emitting galaxy (SMG; $f_{850{\mu}m}\simgt4$~mJy)
  population, and use measured host-galaxy masses to explore their
  evolutionary status. We employ the well-established virial black-hole
  mass estimator to ``weigh'' the black holes of a sample of
  $z\approx$~2 SMGs which exhibit broad H$\alpha$ or H$\beta$
  emission. We find that the average black-hole mass and Eddington
  ratio ($\eta=L_{\rm bol}/L_{\rm Edd}$) of the lower-luminosity
  broad-line SMGs \hbox{($L_{\rm X}\approx10^{44}$~erg~s$^{-1}$)} are
  log($M_{\rm BH}/M_{\odot}$)~$\approx$~8.0 and $\eta\approx$~0.2,
  respectively; by comparison, X-ray luminous broad-line SMGs
  \hbox{($L_{\rm X}\approx10^{45}$~erg~s$^{-1}$)} have log($M_{\rm
  BH}/M_{\odot}$)~$\approx$~8.4 and $\eta\approx$~0.6.  The
  lower-luminosity broad-line SMGs lie in the same location of the
  $L_{\rm X}$--$L_{\rm FIR}$ plane as more typical SMGs hosting X-ray
  obscured active galactic nuclei and may be intrinsically similar
  systems, but orientated so that the rest-frame optical nucleus is
  visible. Under this hypothesis, we conclude that SMGs host black
  holes with log($M_{\rm BH}/M_{\odot}$)~$\approx$~7.8; we find
  supporting evidence from observations of local ultra-luminous
  infrared galaxies.  Combining these black-hole mass constraints with
  measured host-galaxy masses, we find that the black holes in SMGs are
  $\simgt$~3 times smaller than those found in comparably massive
  normal galaxies in the local Universe, albeit with considerable
  uncertainty, and $\simgt$~10 times smaller than those predicted for
  $z\approx$~2 luminous quasars and radio galaxies. These results imply
  that the growth of the black hole lags that of the host galaxy in
  SMGs, in stark contrast with that previously suggested for radio
  galaxies and luminous quasars at $z\approx$~2. On the basis of
  current host-galaxy mass constraints, we show that SMGs and their
  descendants cannot lie significantly above the locally defined
  $M_{\rm BH}$--$M_{\rm GAL}$ relationship. We argue that the black
  holes in the $z\approx$~0 descendents of SMGs will have log($M_{\rm
  BH}/M_{\odot}$)~$\approx$~8.6, indicating that they only need to grow
  by a factor of $\approx$~6 by the present day. We show that this
  amount of black-hole growth can be achieved within current estimates
  for the submm-bright lifetime of SMGs, provided that the black holes
  can grow at rates close to the Eddington limit.

\end{abstract}


%%% MAIN BODY OF TEXT GOES HERE. CONSULT "INSTRUCTIONS FOR AUTHORS USING
%%% LATEX2E MARKUP", SECTIONS 2.3-2.6 FOR HELP WITH EQUATIONS, FIGURES,
%%% AND TABLES.

%\section{}   %%% Top level section head (remove "%" symbol)
%\subsection{}   %%% Second level section head (remove "%" symbol)
%\subsubsection{}   %%% Lowest level section head (remove "%" symbol)
%\section*{}    %%% Unnumbered top level section head (remove "%" symbol)
%\subsection*{}   %%% Unnumbered second level section head (remove "%" symbol)

%
%%%%%%%%%%%%%%%%%%%%%%%%%%%%%%%%%%%%%%%%%%%%%%%%%%%%%%%%%%%%%%%%%%%%%%
\section{Introduction}
%%%%%%%%%%%%%%%%%%%%%%%%%%%%%%%%%%%%%%%%%%%%%%%%%%%%%%%%%%%%%%%%%%%%%%
%

One of the most outstanding astronomical discoveries of the last decade
is the finding that every nearby massive galaxy harbors a central black
hole with a mass directly proportional to that of the galaxy spheroid
(e.g.,\ Magorrian \etal 1998; Ferrarese \& Merritt 2000; Gebhardt \etal
2000; Tremaine \etal 2002; Marconi \& Hunt 2003; H{\"a}ring \& Rix
2004). This landmark result implies that all massive galaxies have
hosted Active Galactic Nuclei (AGN) activity over the past
$\approx$~13~Gyrs (e.g.,\ Soltan 1982; Yu \& Tremaine 2002; Marconi
\etal 2004) and suggests that galaxies and their black holes might have
grown concordantly, despite nine orders of magnitude difference in size
scale (e.g.,\ Silk \& Rees 1998; Fabian 1999; King 2003; Wyithe \& Loeb
2003; Di Matteo \etal 2005; Bower \etal 2006; Croton 2006; Hopkins
\etal 2006b).

Constraining the relative growth of the black hole and spheroid in the
high-redshift progenitors of today's massive galaxies remains a
fundamental goal of cosmology which can shed light on the connection
between AGN activity and star formation, and constrain galaxy formation
models. A number of studies have searched for evolution in the
black-hole--spheroid mass relationship at $z\simgt$~0.5 by ``weighing''
the black holes and host galaxies of luminous AGN populations such as
quasars and radio galaxies (e.g.,\ Shields \etal 2003, 2006; Treu \etal
2004, 2007; Adelberger \& Steidel 2005; McLure \etal 2006; Peng \etal
2006). These studies have typically concluded that the host galaxies of
massive black holes ($\simgt10^{8}$~$M_{\odot}$) at $z\approx$~2 were
up-to $\approx$~6 times less massive than those found for comparably
massive black holes in the local Universe, suggesting that the black
hole may grow before the host galaxy. However, by focusing on luminous
AGNs, these studies are biased toward selecting the most massive and
rapidly growing black holes and could be biased toward evolved
extragalactic source populations (e.g.,\ Lauer \etal 2007). These
studies may therefore miss the early growth phase of today's massive
galaxies where the black-hole--spheroid mass ratio may be
different. This is an important consideration when attempting to model
and understand how black holes and their host galaxies grow.

Arguably, the best candidates for studying the early growth phase of
today's massive galaxies are sub-millimeter (submm) emitting galaxies
(SMGs), such as those identified in deep SCUBA surveys (e.g.,\ Smail
\etal 1997, 2002; Barger \etal 1998; Hughes \etal 1998; Scott \etal
2002; Webb \etal 2003; Borys \etal 2003; Coppin \etal 2006). After
intensive multi-wavelength follow-up observations, it is clear that
bright SMGs ($f_{\rm 850\mu m}\simgt4$~mJy) are ultra-luminous
($\simgt10^{12}$~$L_{\odot}$), gas-rich, massive galaxies at
$z\approx$~2 that are undergoing intense bursts of star formation
(e.g.,\ Swinbank \etal 2004, 2006; Chapman \etal 2005; Greve \etal
2005; Hainline \etal 2006; Kov{\'a}cs \etal 2006; Tacconi \etal 2006);
their star-formation rates are large enough to form a massive galaxy of
$10^{11}$~$M_{\odot}$ in just (1--2)$\times10^{8}$~yrs. The catalyst
for this intense activity is probably galaxy major mergers (e.g.,\
Smail \etal 1998; Chapman \etal 2003; Conselice \etal 2003), which are
predicted by hydrodynamical simulations to provide an effective
mechanism to transport material toward the central region of the galaxy
and trigger nuclear star formation and AGN activity (e.g.,\ Di Matteo
\etal 2005; Springel \etal 2005).

Ultra-deep {\it Chandra} observations (the 2~Ms {\it Chandra} Deep
Field-North survey; Alexander \etal 2003a) have shown that at least
$\approx$~28--50\% of the bright SMG population host heavily obscured
AGNs (e.g.,\ Alexander \etal 2003b, 2005a,b), implying joint growth of
the galaxy and black hole (see also Ivison \etal 2002; Wang \etal 2004;
Pope \etal 2006).\footnote{The sensitivity of instruments prevents good
statistics on the prevalence of AGN activity in the faint SMG
population ($f_{\rm 850\mu m}\simlt4$~mJy).} The energetics of the AGNs
measured at X-ray and mid-infrared (mid-IR) wavelengths are (typically)
too low to explain the huge bolometric output of these objects ($L_{\rm
bol}\approx10^{12}$--$10^{13}$~$L_{\odot}$; e.g.,\ Ivison \etal 2004;
Alexander \etal 2003b, 2005b; Lutz \etal 2005), which appears to be
dominated by star-formation processes (as shown by the strong
star-formation signatures in the mid-IR spectra; e.g.,\
Men{\'e}ndez-Delmestre \etal 2007; Valiante \etal 2007; Pope \etal
2008).\footnote{We note that in this respect SMGs are unlike the local
ULIRG Arp~220, which has ambiguous AGN and star-formation signatures
(e.g.,\ Haas et~al. 2001; Downes \& Eckart 2007).} However, the large
AGN fraction implies that the black holes are growing almost
continously throughout periods of vigorous star formation. This
continuous black-hole growth suggests an abundance of available fuel,
and hints that the accretion may be occuring at the Eddington limit
(i.e.,\ exponential black-hole growth; e.g.,\ Rees 1984). Although
hypothetical, this picture is in good agreement with models for the
growth of black holes in SMG-like systems (e.g.,\ Archibald \etal 2002;
Di Matteo \etal 2005; King 2005; Granato \etal 2006; c.f.,\ Chakrabarti
\etal 2007).

Utilising deep rest-frame UV--near-IR observations, Borys \etal (2005)
estimated the stellar masses of the X-ray obscured SMGs identified in
Alexander \etal (2005a). They suggested that the Eddington-limited
black-hole masses of these objects ($M_{\rm BH,
Edd}\approx10^{7}$~$M_{\odot}$) are $\approx$~50 times smaller than the
black holes hosted by comparably massive normal galaxies ($M_{\rm
GAL}\approx2.5\times10^{11}$~$M_{\odot}$) in the local Universe. This
result suggests that either (1) the growth of the black hole lags that
of the host galaxy in SMGs, or (2) the black holes in SMGs are
accreting at sub-Eddington rates. The former is in stark contrast with
the growth of black holes determined for high-redshift quasars and
radio galaxies (e.g.,\ McLure \etal 2006; Peng \etal 2006) while the
latter is in conflict with that predicted by most models for the growth
of SMG-like systems with an abundance of available fuel (e.g.,\
Archibald \etal 2002; Granato \etal 2006). More direct constraints on
the mases and Eddington ratios ($\eta=L_{\rm bol}/L_{\rm Edd}$; i.e.,\
the fraction of the Eddington limit) of the black holes in SMGs are
required to distinguish between these different scenarios. These
results can then be used to place constraints on the relative
black-hole--galaxy growth of this key high-redshift population,
including relating SMGs to other high-redshift populations such as
quasars and radio galaxies.

In this paper we employ the well-established virial black-hole mass
estimator (e.g.,\ Wandel \etal 1999; Kaspi \etal 2000) to directly
determine black-hole masses ($M_{\rm BH}$) and $\eta$ for the small
subset of SMGs that have broad emission lines. The virial black-hole
mass estimator provides the most direct and accessible method currently
available to determine black-hole masses at high redshift, and this is
the first study that has used this technique to estimate $M_{\rm BH}$
for the cosmologically important SMG population. We use these results
to estimate the black-hole properties of the X-ray obscured SMGs
identified in Alexander \etal (2005a,b).  Observations of nearby
Ultra-Luminous IR Galaxies ($L_{\rm IR}\simgt10^{12}$~$L_{\odot}$;
ULIRGs; e.g.,\ Sanders \& Mirabel 1996) hosting obscured AGNs are also
used to determine the Eddington ratios and intrinsic AGN properties of
these $z\approx$~0 analogs to the X-ray obscured SMG population and, by
inference, place indirect constraints on the properties of typical
SMGs. We have adopted $H_{0}=71$~km~s$^{-1}$~Mpc$^{-1}$, $\Omega_{\rm
M}=0.27$, and $\Omega_{\Lambda}=0.73$ throughout.

%
%%%%%%%%%%%%%%%%%%%%%%%%%%%%%%%%%%%%%%%%%%%%%%%%%%%%%%%%%%%%%%%%%%%%%%
\section{Samples and Method}
%%%%%%%%%%%%%%%%%%%%%%%%%%%%%%%%%%%%%%%%%%%%%%%%%%%%%%%%%%%%%%%%%%%%%%
%

\subsection{Samples}

We have defined several samples to use in this study, which we briefly
describe below. We have used two SMG samples, a broad-line SMG sample,
to enable us to directly estimate $M_{\rm B H}$ and $\eta$, and an
X-ray obscured SMG sample, which is more representative of the overall
SMG population than the broad-line SMG sample. The broad-line SMG
sample is further broken down into ``lower-luminosity broad-line
SMGs'', which includes objects with $L_{\rm
X}\approx10^{44}$~erg~s$^{-1}$, and ``X-ray luminous broad-line SMGs'',
which includes objects with $L_{\rm X}\approx10^{45}$~erg~s$^{-1}$; see
\S3.1 for details of the sample selection.  For the X-ray obscured SMG
sample we focus on the $z>1.8$ SMGs that host X-ray identified AGNs in
Alexander \etal (2005a,b), to probe the peak epoch of SMG activity
(e.g.,\ $z\approx$~2.2; Chapman \etal 2005); we thus removed three
$z\simlt1$ SMGs, which make up the low-redshift tail of the SMG
population. The average redshift of the $z>1.8$ X-ray obscured SMG
sample ($z=2.29\pm0.32$) is consistent with that of the broad-line SMG
sample ($z=2.36\pm0.26$).

We have also defined two samples of nearby ULIRGs to provide more
direct insight into the black-hole masses and AGNs of dust-obscured
objects with similar bolometric luminosities to the SMG population. The
first sample comprises ULIRGs hosting obscured AGN activity for which
we were able to estimate $M_{\rm BH}$ and $\eta$ from the presence of
broad near-IR emission lines; see \S4.1 for details of the sample
selection. The second sample is distance limited and comprises all of
the ULIRGs hosting AGN activity that lie within 200~Mpc. The X-ray and
far-IR data of this sample is used to assess whether the intrinsic
properties of the AGNs in the X-ray obscured SMG sample have been
underestimated, which would lead to underestimates in $M_{\rm BH}$; see
\S4.2 for details of the sample selection.

%\newpage
\subsection{Method}

Black-hole masses have been directly measured from the velocity
dispersion of stars/gas in the nuclear regions of galaxies in the local
Universe (e.g.,\ Kormendy \& Richstone 1995; Sarzi \etal 2001; Gebhardt
\etal 2003; Pinkney \etal 2003). Black holes cannot be ``weighed''
using the same techniques at high redshift due to poorer spatial
resolution and lower signal-to-noise ratio data. However, the
well-established virial black-hole mass estimator, which works on the
assumption that the broad-line region (BLR) in AGNs is under the
gravitational influence of the black hole (i.e.,\ $V^2=GM/r$; see
Peterson \& Wandel 1999, 2000 for evidence), provides an apparently
reliable, if indirect, measurement of $M_{\rm BH}$ for more distant
AGNs (e.g.,\ Wandel \etal 1999; Kaspi \etal 2000; McLure \& Dunlop
2002; Vestergaard 2002; Peterson \etal 2004; Vestergaard \& Peterson
2006). Using this technique, the masses of black holes in quasars up to
$z\approx$~6.4 have been estimated (e.g.,\ Willott \etal 2003; McLure
\& Dunlop 2004; Vestergaard 2004). The calibration of this technique is
predominantly restricted to reverberation mapping of nearby AGNs in the
local Universe, however, the first reverberation mapping experiments at
high redshift are already underway (see Kaspi 2007; Kaspi \etal 2007).

The primary aim of this paper is to estimate $M_{\rm BH}$ for the X-ray
obscured SMG sample to determine the relative black-hole--galaxy growth
of SMGs and explore their evolutionary status. The X-ray obscured SMGs
already have host-galaxy mass constraints (e.g.,\ Borys \etal 2005),
and since $\approx$~28--50\% of SMGs host heavily obscured AGNs they
should be representative of the SMG population.  However, the presence
of heavy absorption in these SMGs hides the BLR in the majority of
these systems (e.g.,\ Swinbank \etal 2004; Alexander \etal 2005b;
Chapman \etal 2005), making it difficult to directly use the virial
black-hole mass estimator to determine their black-hole
properties. Fortunately, the identification of broad emission lines
from a small number of SMGs (e.g.,\ Swinbank \etal 2004; Takata \etal
2006) provides the opportunity to estimate the black-hole masses and
Eddington ratios for this subset of the SMG population, providing a
lever to determine constraints for the, more typical, X-ray obscured
SMGs. For example, taking the canonical approach that the relationship
between broad-line AGNs and obscured AGNs is the orientation of an
optically and geometrically thick torus (i.e.,\ the unified AGN model;
e.g.,\ Antonucci 1993), the average Eddington ratios and black-hole
masses of the X-ray obscured SMGs should be comparable to the
broad-line SMGs. In this scenario we would calculate the black-hole
masses of the SMGs as

\begin{equation}
M_{\rm BH (XO-SMG)}= {{M_{\rm BH,Edd (XO-SMG)}}\over{\eta_{\rm (BL-SMG)}}}
\end{equation}

Where $M_{\rm BH (XO-SMG)}$ and $M_{\rm BH,Edd (XO-SMG)}$ are the
black-hole mass and X-ray derived Eddington-limited black-hole mass for
the X-ray obscured SMGs, and $\eta_{\rm (BL-SMG)}$ is the Eddington
ratio for the broad-line SMGs, which is defined as

\begin{equation}
\eta= {\kappa {L_{\rm X}}\over{L_{\rm Edd}}}
\end{equation}

Where $\kappa_{\rm 2-10 keV}$ is the X-ray-to-bolometric luminosity
conversion.  Our overall approach is similar in essence to that adopted
by McLure \etal (2006) when exploring the black-hole--galaxy mass ratio
for radio galaxies. We investigate $M_{\rm BH}$, $\eta$, and the
black-hole--galaxy mass ratio for the SMGs in \S3.

However, it is also possible that the relationship between the
broad-line SMGs and the X-ray obscured SMGs is not the orientation of
an obscuring torus toward the line of sight. For example, the
broad-line SMGs might represent a more evolved stage in the evolution
of the SMG population or they may be undergoing a period of increased
mass accretion that makes the BLR temporarily visible.  Since an
optically bright broad-line AGN is often considered to be at a later
phase in the evolution of massive galaxies than an X-ray obscured AGN
(e.g.,\ Sanders \etal 1988; Granato \etal 2006; Hopkins \etal 2006a;
Chakrabarti \etal 2007), in the evolutionary scenario, the black-hole
masses of the broad-line SMGs will provide an upper limit to the
black-hole masses of the X-ray obscured SMGs. By comparison, in the
increased mass accretion scenario, the black-hole masses of the X-ray
obscured SMGs would be the same as the broad-line SMGs but the
Eddington ratios would be lower. Unless we can rule out these
alternative scenarios, our determination of $M_{\rm BH}$ for the X-ray
obscured SMGs will be subject to larger uncertainties.  However,
although the presence of absorption in the X-ray obscured SMGs hinders
direct black-hole mass measurements, the identification of broad
Pa$\alpha$ in a small number of nearby obscured ULIRGs provides the
potential to gain insight into the Eddington ratios for obscured AGNs
similar to those found in the X-ray obscured SMGs (Veilleux \etal 1997,
1999), and hence distinguish between these different scenarios.  The
broad Pa$\alpha$ emission from these obscured ULIRGs is thought to be
observed through the atmosphere of a dusty obscuring torus, which is
assumed to be optically thin at near-IR wavelengths. The similarity
between the bolometric luminosities, spectral energy distributions
(SEDs), and morphological properties of SMGs and nearby ULIRGs suggests
that distant SMGs may be ``scaled-up'' versions of ULIRGs, providing
validation for this approach [e.g.,\ mid-IR spectra
(Men{\'e}ndez-Delmestre \etal 2007; Valiante \etal 2007), submm
dust/gas emission and extent (Greve \etal 2005, 2006; see Table~2 in
Tacconi \etal 2006), major mergers (Chapman \etal 2003; Conselice \etal
2003)]. We determine $M_{\rm BH}$, $\eta$, and the black-hole--galaxy
mass ratio for the nearby ULIRGs in \S4.

%
%%%%%%%%%%%%%%%%%%%%%%%%%%%%%%%%%%%%%%%%%%%%%%%%%%%%%%%%%%%%%%%%%%%%%%
\section{Weighing the Black Holes of $z\approx$~2 Submillimeter Galaxies Hosting AGN Activity}
%%%%%%%%%%%%%%%%%%%%%%%%%%%%%%%%%%%%%%%%%%%%%%%%%%%%%%%%%%%%%%%%%%%%%%
%

In this section we determine $M_{\rm BH}$ for the broad-line SMGs using
the virial black-hole mass estimator of Greene \& Ho (2005), which
calculates black-hole masses solely from the properties of the
H$\alpha$ or H$\beta$ emission line, and reduces potential
uncertainties on the luminosity of the AGN (e.g.,\ contaminating
emission from the host galaxy or an accretion-related jet) when
compared to other estimators. We estimate Eddington ratios for these
sources using the black-hole mass and the X-ray luminosity, as a proxy
for the mass accretion rate. We then compare the properties of the
X-ray obscured SMGs to the broad-line SMGs and use these results to
constrain $M_{\rm BH}$ and the black-hole--galaxy mass ratio for the
more typical X-ray obscured SMGs.

\subsection{The selection of broad-line SMGs}

The broad-line SMG sample is selected from the redshift survey of
submm-emitting galaxies by Chapman \etal (2005), the same original
source as the X-ray obscured SMG sample (Alexander \etal 2005a,b). The
initial sample comprises 12 SMGs with broad (FWHM$>1000$~km~s$^{-1}$)
H$\alpha$ or H$\beta$ emission from Swinbank \etal (2004), Takata \etal
(2006), and new Keck NIRSPEC data (K.~Men{\'e}ndez-Delmestre et al., in
prep). We removed six objects where the [OIII]$\lambda$5007
emission-line profile was found to be similar to the H$\alpha$/H$\beta$
emission-line profile. [OIII]$\lambda$5007 is a forbidden line that is
not intrinsically broad (typically FWHM$\simlt600$~km~s$^{-1}$),
suggesting that the broadening of the H$\alpha$/H$\beta$ emission in
these six objects might not be intrinsic to the BLR and could, instead,
be due to external kinematics (e.g.,\ outflows and multiple components
within galaxy mergers; e.g.,\ Heckman \etal 1981; Whittle 1985,
1988). However, before applying the virial black-hole mass estimator to
the remaining six objects in the broad-line SMG sample it is necessary
to verify that the BLR emission is not dominated by scattered light (as
would be expected if the BLR was completely obscured; e.g.,\ Antonucci
\& Miller 1985; Young \etal 1996) and to correct the broad-line
emission for extinction.

The detection of submm emission from the broad-line SMGs implies the
presence of dust, which might obscure the BLR in some cases. To test if
the BLR emission is seen directly (rather than in scattered light) we
compared the broad-line luminosity to the rest-frame 2--10~keV
luminosity; see Table~1 for the source properties. We found that the
broad-line SMGs have stronger H$\alpha$ emission than expected from the
local H$\alpha$--X-ray luminosity relationship of Ward \etal (1988),
indicating that the BLR is brighter than found in typical broad-line
AGNs, and demonstrating that the emission is being seen
directly. Indeed, the H$\alpha$--X-ray luminosity ratios of the
broad-line SMGs are similar to those of nearby ULIRGs ($L_{\rm
X}/L_{H\alpha}\approx3$, a factor $\approx$~6 higher than the average
in Ward \etal 1988; e.g.,\ Imanishi \& Terashima 2004).

We estimated the extinction to the BLR for each source on the basis of
the broad H$\alpha$/H$\beta$ emission-line ratio, following the
prescription of Ward \etal (1987) for an intrinsic ratio of
H${\alpha}$/H${\beta}=$~3.1 (i.e.,\ Case B recombination; Osterbrock
1989) and the reddening curve of Calzetti \etal (2000). Extinction
corrections are rarely applied to the broad-line luminosities of AGNs
used in studies estimating virial black-hole masses, even when
rest-frame ultra-violet (UV) emission lines are used. However, even in
our potentially dust obscured sample of broad-line AGNs we did not find
evidence for large amounts of extinction (the mean extinction
correction was $A_V\approx$~1.2~mags). The comparatively low level of
extinction suffered by the nucleus when compared to the high extinction
suffered by the star-forming regions in SMGs (e.g.,\ Swinbank \etal
2004; Takata \etal 2006) could be due to the dust tracing the extended
star-forming regions rather than the nucleus; we note that by selecting
broad-line AGNs our sample will also be biased toward objects with low
levels of nuclear extinction. The intrinsic broad
H${\alpha}$/H${\beta}$ emission-line ratio in many broad-line AGNs is
larger than the theoretical Case B limit (e.g.,\ Adams \& Weedman 1975;
Osterbrock 1977), which would give slightly smaller black-hole masses
for our broad-line SMGs.

Two of the broad-line SMGs did not have H$\alpha$ emission-line
constraints; see Table~1. However, since these two objects are the
optically brightest $z\approx$~2 SMGs in the Chapman \etal (2005)
sample and have rest-frame UV and/or lower-order Balmer broad emission
lines in their spectra (Chapman \etal 2005; Takata \etal 2006), the
extinction to their BLRs are likely to be small. Furthermore, since the
average broad H$\alpha$/H$\beta$ emission-line ratio of the AGNs used
in the Greene \& Ho (2005) virial black-hole mass estimation is larger
than the expected intrinsic ratio (the average broad H$\alpha$/H$\beta$
ratio of the sample is 3.5), this effectively allows for the inclusion
of small amounts of reddening, and hence we adopted zero reddening for
these two SMGs.

The basic properties of the broad-line SMG sample are given in
Table~1. Although this sample is not complete, it is representative of
SMGs spanning a wide range in X-ray luminosity. We make a distinction
between the broad-line SMGs studied here and X-ray selected quasars
detected at submm wavelengths (e.g., Page \etal 2001; 2004; Stevens
\etal 2005), which have smaller space densities than the SMG
populations studied here; however, see \S5.2 and Coppin \etal (2008)
for possible overlap between these source populations.

%
%%%%%%%%%%%%%%%%%%%%%%%%%%%%%%%%%%%%%%%%%%%%%%%%%%%%%%%%%%%%%%%%%%%%%%
% TABLE 1
%%%%%%%%%%%%%%%%%%%%%%%%%%%%%%%%%%%%%%%%%%%%%%%%%%%%%%%%%%%%%%%%%%%%%%
%

\begin{table}[!t]
 \centering
  \caption{Broad-line SMGs}
  \begin{tabular}{@{}lccccccccc@{}}
  \hline
 Name        &     &   log($L_{\rm H\alpha}$)$^{\rm a}$ & FWHM$_{H\alpha}$
 & log($M_{\rm BH}$) & log($L_{\rm X}$)$^{\rm b}$ & \\
 (SMMJ)    & $z$ &   (erg~s$^{-1}$) & (km~$s^{-1}$)    & ($M_{\odot}$)         & (erg~s$^{-1}$) & Refs\\
 \hline
123635.5+621424  &    2.015  &     43.4  &    $1600\pm200$  &         7.3 & 43.8 & 1,2\\
123716.0+620323  &    2.053  &     44.1  &    $2400\pm500$  &         8.2 & 44.1 & 2,3\\
131215.2+423900  &    2.555  &     43.7$^{\rm c}$  &         $2500\pm500$$^{\rm c}$  &     7.9 & 44.9 & 4,5\\
131222.3+423814  &    2.560  &     44.3$^{\rm c}$  &         $2600\pm1000$$^{\rm c}$ &     8.3 & 44.7 & 4,5\\
163655.8+405914   &    2.592  &     44.4  &    $3000\pm400$  &         8.5  & 45.0 & 1,6\\
163706.5+405313  &    2.375  &     43.3  &    $3300\pm1000$  &         7.9 & $<44.0$ & 1,6\\
\hline
\vspace{0.01in}
\end{tabular}
\begin{minipage}{85mm}
$^a$~The H$\alpha$ luminosity has been corrected for extinction,
determined from the broad-line H$\alpha$/H$\beta$ flux ratio, except
for the two 13~hr sources for which only H$\beta$ data is available
(these two sources are optically bright and probably
do not suffer significant extinction; see \S3.1).\\ 
$^b$~Rest-frame 2--10~keV band, corrected for absorption;
see \S3.3.\\ 
$^c$~Line width is for H$\beta$. The H$\alpha$ 
luminosity is calculated from the
H$\beta$ luminosity, assuming an intrinsic H$\alpha$/H$\beta=3.1$.\\
{\sc References:} --
(1) Swinbank \etal (2004); (2) Alexander \etal (2003a); (3)
K.~Men{\'e}ndez-Delmestre \etal, in prep; (4) Takata \etal (2006); (5)
Mushotzky \etal (2000); (6) Manners \etal (2003).\\
\end{minipage}
\end{table}

\vspace{0.5in}
\subsection{Weighing the black holes in broad-line SMGs}

%
% Figure 1: H-alpha properties
%
\begin{figure}[!t]
\plotone{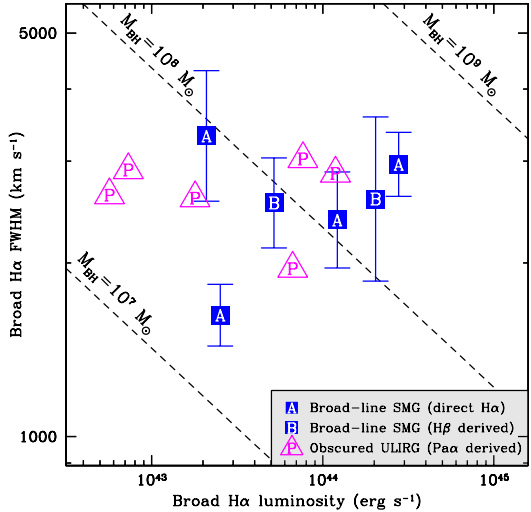}
\caption{H$\alpha$ properties of the broad-line SMGs
and obscured ULIRGs. Sources indicated with letters
have had their H$\alpha$ properties determined from their
extinction-corrected H$\alpha$ emission (``A''), derived from their
H$\beta$ properties for an intrinsic H$\alpha$/H$\beta$ ratio of 3.1
(``B''; see \S3.1), or derived from their Pa$\alpha$ properties for an
intrinsic H$\alpha$/Pa$\alpha$ ratio of 8.6 (``P''; see \S4.1).
The dashed lines show the relationship between the
black-hole mass ($M_{\rm BH}$) and the broad H$\alpha$ emission-line
properties on the basis of the virial black-hole mass estimator of
Greene \& Ho (2005). Taken at face value the data suggest that
broad-line SMGs have a range of black-hole masses with a mean of
log($M_{\rm BH}/M_{\odot}$)~$\approx$~8.0; the obscured ULIRGs have a
mean black-hole mass of log($M_{\rm BH}/M_{\odot}$)~$\approx$~7.8. However,
if it is assumed that the BLR gas in the broad-line SMGs is
distributed in a disk that is seen close to pole on (rather than the
BLR gas being isotropically distributed and in random motion) then the
average black-hole mass will be $\approx$~2.7 times larger, on the
basis of the model described in the text (see \S3.2).} 
\end{figure}

%
% Figure 2: Mass accretion properties
%
\begin{figure*}[!t]
\centerline{\includegraphics[angle=0,width=12cm]{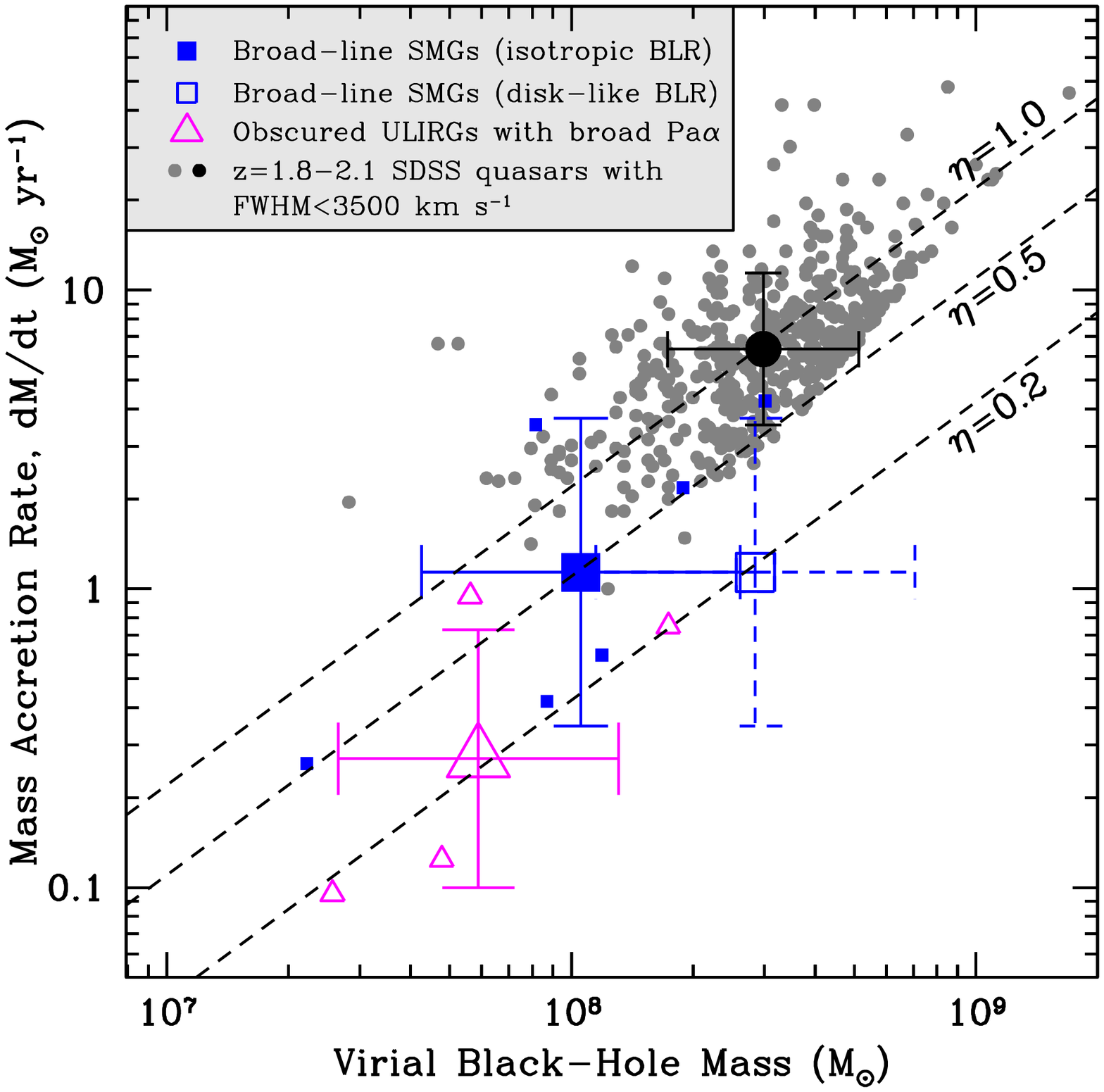}}
\figcaption{Mass accretion rate versus black-hole mass of the
broad-line SMG sample, a sample of nearby ULIRGs hosting obscured AGNs
but with detected broad Pa$\alpha$ emission, and $z=$~1.8--2.1 SDSS
quasars with comparatively narrow emission lines
(FWHM$<3500$~km~s$^{-1}$; data taken from McLure \& Dunlop 2004). The
error bars correspond to $1\sigma$ uncertainties on the average
properties.  The dashed lines indicate Eddington ratios ($\eta=L_{\rm
bol}/L_{\rm Edd}$; i.e.,\ the fraction of the Eddington limit) of
$\eta=0.2$, $\eta=0.5$, and $\eta=1.0$. The average Eddington ratio of
the broad-line SMGs is $\eta\approx0.5$, assuming that the BLR gas is
isotropically distributed and in random motion, or $\eta\approx0.2$,
assuming that the BLR gas is distributed in a disk that is observed
close to pole on (see \S3.2); this is compared to the average for the
SDSS quasars with comparatively narrow emission lines ($\eta\approx1.0$
for BLR gas in random motion). The average Eddington ratio of the
nearby ULIRGs hosting obscured AGNs, where inclination-angle effects
will not be as important, is $\eta\approx0.2$ (see \S4.1).}
\vspace{0.15in}
\end{figure*}

Fig.~1 shows the emission-line properties of the broad-line SMG
sample. Using the Greene \& Ho (2005) virial black-hole mass estimator,
the black-hole masses of the broad-line SMG sample are log($M_{\rm
BH}/M_{\odot}$)~$\approx$~7.3--8.5 and the mean is log($M_{\rm
BH}/M_{\odot}$)~=~$8.0\pm0.4$. By comparison, optically selected
$z\approx$~2 quasars in the Sloan Digital Sky Survey (SDSS) have
black-hole masses about an order of magnitude higher [log($M_{\rm
BH}/M_{\odot}$)~=~$8.9\pm0.3$; see Fig.~1 in McLure \& Dunlop
2004]. This difference is largely due to selection effects as the SDSS
quasars are luminous at rest-frame UV wavelengths ($i<19.1$; Schneider
\etal 2003), and are therefore biased toward the largest black holes
(i.e.,\ rarer objects), while all but two of the broad-line SMGs have
$i>19.1$. However, the comparative narrowness of the broad lines in
these SMGs when compared to the typical quasar population
($\approx$~2500~km~s$^{-1}$ versus $\approx$~5000~km~s$^{-1}$) suggests
relatively modest black-hole masses. For comparison, the mean
black-hole mass of $z=$~1.8--2.1 SDSS quasars with
FWHM~$<3500$~km~s$^{-1}$ is log($M_{\rm BH}/M_{\odot}$)~=~$8.5\pm0.2$
(McLure \& Dunlop 2004).

The broad-line SMG black-hole mass constraints are calculated for the
scenario where the BLR gas is isotropically distributed and in random
motion within the gravitational potential of the black hole, as
typically assumed. However, if the BLR gas is distributed in a disk
which is observed close to pole on then the broad-line width will only
provide a lower limit to the intrinsic velocity of the BLR (and hence
the black-hole mass; e.g.,\ Jarvis \& McLure 2006). Indeed, the
narrowness of the broad lines indicates that our broad-line SMG sample
may be biased toward objects seen pole on (McLure \& Dunlop 2002;
Collin \etal 2006; Jarvis \& McLure 2006). We can assess the potential
effect of a disk-like BLR by assuming that the average inclination
angle of the BLR can be calculated from the obscured:unobscured AGN
ratio of the SMG population, under the hypothesis of the unified AGN
model.  On the basis of the X-ray spectral analysis of SMGs by
Alexander et al. (2005b), the obscured:unobscured AGN ratio for the
SMGs is $\approx$~5--10, giving a maximum BLR inclination for the
broad-line SMGs of $i<$~25--34$^{\circ}$ and an average BLR inclination
of $i=$~18--25$^{\circ}$ (e.g.,\ Eqn.~1 in Arshakian 2005). Using the
BLR model of McLure \& Dunlop (2002), the average correction to the
black-hole mass for a disk-like BLR would be $\approx$~2.7, giving a
mean mass for the broad-line SMG sample of log($M_{\rm
BH}/M_{\odot}$)~=~$8.4\pm0.4$; see Fig.~1 in McLure \& Dunlop
(2002).\footnote{We highlight here the close similarity between the
estimated average inclination angle of the BLR and that estimated for
the CO emission in the host galaxies of submm-detected quasars
($i\approx13^{\circ}$; Carilli \& Wang 2006).}  There is no conclusive
evidence that the BLR is orientated in a disk in radio-quiet AGNs,
however, for completeness we conservatively consider both average
black-hole masses in these initial analyses.

\subsection{The Eddington ratios of broad-line SMGs}

The Eddington ratio of a black hole is calculated from the black-hole
mass and the mass accretion rate (e.g.,\ Rees 1984) and provides a
measurement of the black hole ``growth time'' (i.e.,\ the time required
for the black hole to double in mass); see Eqn.~2. The determination of
accurate Eddington ratios is challenging since they rely upon excellent
measurements of the black-hole mass, the bolometric correction, and the
conversion between the AGN bolometric luminosity and the mass accretion
rate. However, the primary reason that we calculate Eddington ratios in
this paper is to provide a way to scale between the measured black-hole
masses of the broad-line SMGs and the Eddington-limited black-hole
masses of the X-ray obscured SMGs; see Eqn.~1. The Eddington ratios
derived in this paper should therefore be considered indicative rather
than absolute values (as is generally the case for high-redshift
studies). An advantage of this approach is that it is only important
that the X-ray-to-bolometric luminosity corrections of the broad-line
SMGs and the X-ray obscured SMGs are the same.  This assumption seems
reasonable given that (1) the broad-line SMGs and the X-ray obscured
SMGs have similar X-ray and far-IR luminosities (see \S3.4 and Fig.~3),
and (2) the broad-line SMGs and obscured ULIRGs have similar Eddington
ratios (see \S4).

Here we calculate the Eddington ratio for the broad-line SMGs using the
X-ray luminosity as a proxy for the mass accretion rate. We converted
the X-ray luminosity to the bolometric AGN luminosity using the Elvis
\etal (1994) mean spectral energy distribution ($\kappa_{\rm 2-10
keV}\approx$~35); see Eqn.~2.  We note that using different bolometric
conversions (e.g.,\ Marconi et~al. 2004; Vasudevan \& Fabian 2007) will
increase our estimate of the Eddington ratio by up-to a factor of
$\approx$~2 (up-to $\kappa_{\rm 2-10 keV}\approx$~70 for high
accretion-rate AGNs). However, adopting different bolometric
conversions will not change our determination of the black-hole masses
for the X-ray obscured SMGs; see Eqn.~3 \& \S3.4. The X-ray
luminosities of the six broad-line SMGs were determined using {\it
Chandra} data from Mushotzky \etal (2000), Manners \etal (2003), and
Alexander \etal (2003a). When appropriate, X-ray absorption corrections
were made following Alexander \etal (2005b); see Fig.~5 and \S4.2.

In Fig.~2 we show the X-ray derived mass accretion rate versus virial
black-hole mass for the broad-line SMG sample; see \S4.1 for a
discussion of the obscured ULIRGs. The most luminous broad-line SMGs
have an average Eddington ratio comparble to those of $z=$~1.8--2.1
SDSS quasars with narrow broad lines ($\eta\simgt0.5$; e.g.,\ McLure \&
Dunlop 2004).  However, an estimate of the average Eddington ratio for
the broad-line SMG sample depends upon the distribution of the gas in
the BLR. If the BLR gas is isotropically distributed and in random
motion, as typically assumed, then the average Eddington ratio is
$\eta=0.5^{+1.0}_{-0.3}$, while the average Eddington ratio drops to
$\eta=0.2^{+0.4}_{-0.1}$ if the BLR gas is distributed in a disk that
is observed close to pole on. The higher value is consistent with
Eddington-limited accretion while the lower value is similar to those
found for the distant broad-line AGN population (e.g.,\ McLure \&
Dunlop 2004; Kollmeier \etal 2006) and would appear to suggest that
broad-line SMGs are not growing their black holes at a significantly
faster rate than typical submm-undetected broad-line AGNs (only
$\approx$~5--15\% of $z\approx$~2 quasars are submm detected; Priddey
\etal 2003; Page \etal 2004).

\subsection{The Black-hole masses of the X-ray obscured SMGs}

%
% Figure 3: X-ray versus far-IR luminosity
%
\begin{figure}[!t]
\plotone{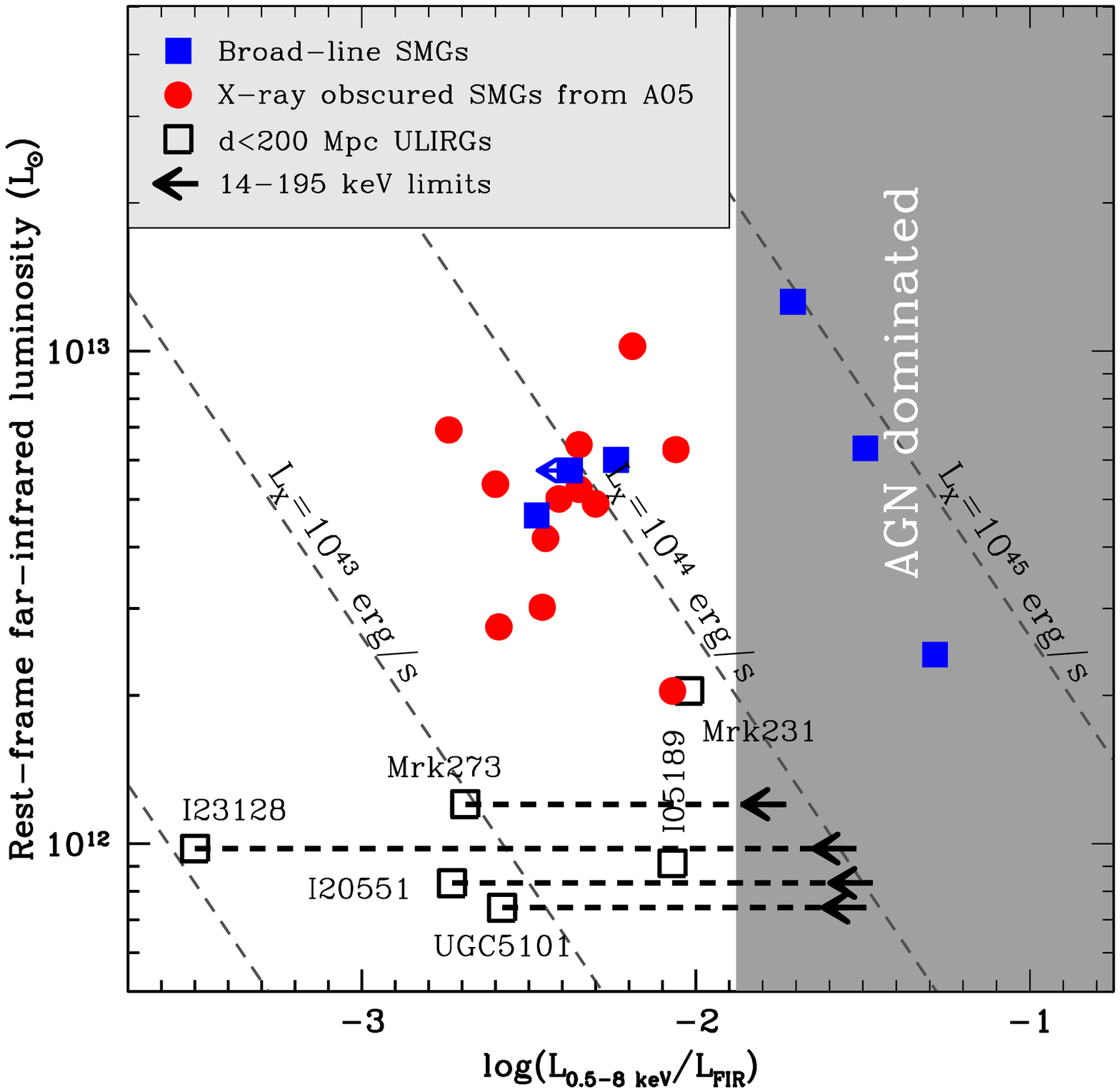}
\caption{Rest-frame far-IR luminosity versus X-ray-to-far-IR luminosity
ratio for the X-ray obscured SMGs, the broad-line SMGs, and the
\hbox{$d<200$~Mpc} ULIRGs; the X-ray luminosities are corrected for
absorption, where necessary. The arrows indicate the 14--195~keV limits
for the \hbox{$d<200$~Mpc} ULIRGs from {\it Swift}-BAT (Markwardt \etal
2005); both Mrk~231 and I05189-2524 are detected at high X-ray energies
(20--50~keV) using {\it BeppoSAX}-PDS (Dadina 2007). The far-IR
luminosities for the X-ray obscured SMG and the broad-line SMGs are
calculated from the radio luminosity assuming the radio-to-far-IR
relationship (the former are taken from Alexander \etal 2005b), and the
far-IR luminosities for the $d<200$~Mpc ULIRGs are taken from Sanders
\etal (2003). The diagonal dashed lines indicate different X-ray
luminosities and the shaded region shows where AGN-dominated sources
will lie on the basis of the Elvis \etal (1994) spectral energy
distribution (taken from Fig.~8 of Alexander \etal 2005b). The three
lower-luminosity broad-line SMGs lie in the same location of the
$L_{\rm X}$--$L_{\rm FIR}$ plane as the X-ray obscured SMGs, however,
the other three broad-line SMGs have X-ray luminosities $\approx$~10
times higher than the broad-line SMGs and are not typical of the SMG
population (these X-ray luminous broad-line SMGs are probably AGN
dominated systems).}
\end{figure}

%
% Figure 4: M-sigma results
%
\begin{figure*}[!t]
\centerline{\includegraphics[angle=0,width=12cm]{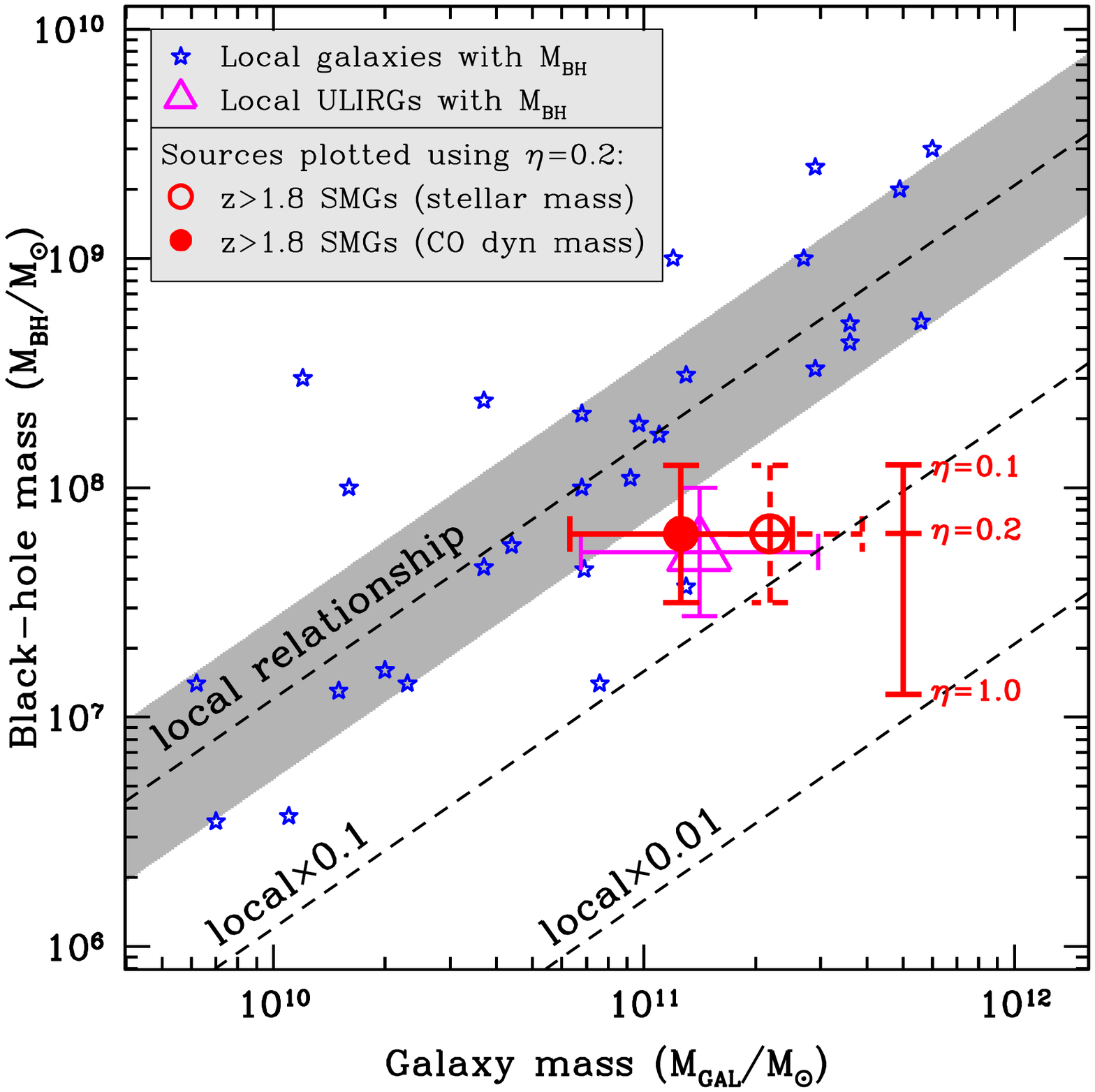}}
\figcaption{Black-hole versus galaxy mass. We plot the estimated
average properties of the X-ray obscured SMGs for different host-galaxy
mass measurements (CO dynamics; Greve \etal 2005; stellar masses; Borys
\etal 2005), assuming $\eta=0.2$. We also plot the average properties
of the obscured ULIRGs with broad Pa$\alpha$ emission (see \S4.1) and
show the properties of well-studied nearby galaxies (taken from
H{\"a}ring \& Rix 2004). The error bars correspond to $1\sigma$
uncertainties on the average properties. The shaded region indicates
the black hole--spheroid mass relationship from H{\"a}ring \& Rix
(2004) and the dashed lines indicate different ratios of this
relationship. The solid bar indicates the black-hole mass constraints
for the X-ray obscured SMGs for different Eddington ratios. For
$\eta=0.2$, the SMGs lie $\approx$~3--5 times below this relationship,
depending upon whether the average CO dynamical mass or stellar mass
constraints are used; these results are qualitatively consistent with
those found for the nearby obscured ULIRGs.}
\vspace{0.15in}
\end{figure*}

Due to the presence of heavy obscuration in the majority of the SMG
population, we are unable to use the virial black-hole mass estimator
to constrain directly the black-hole masses of the X-ray obscured
SMGs. However, if the X-ray obscured SMGs are related to the broad-line
SMGs either via the orientation of an obscuring torus (e.g.,\ Antonucci
1993) or in terms of an evolutionary scenario (e.g.,\ Sanders \etal
1988), then we can use the black-hole masses and Eddington ratios of
the broad-line SMGs to constrain the black-hole masses of the X-ray
obscured SMGs; see \S2 for discussion of these scenarios.

For example, applying the Eddington ratios of the broad-line SMGs
($\eta=$~0.2--0.5) to the average Eddington-limited black-hole mass of
the $z>1.8$ X-ray obscured SMG sample [log($M_{\rm
BH,Edd}$/$M_{\odot}$)~$\approx7.1$; Alexander \etal 2005a] gives
\hbox{log($M_{\rm BH}$/$M_{\odot}$)~$\approx$~7.4--7.8}; see Eqn.~1.
These average black-hole masses are $\approx$~4 times smaller than
those estimated for the broad-line SMGs, which is due to the difference
in average X-ray luminosity between the X-ray obscured SMGs and the
broad-line SMGs (mean 0.5--8.0~keV luminosities of
$\approx8\times10^{43}$~erg~s$^{-1}$ and
$\approx3\times10^{44}$~erg~s$^{-1}$, respectively); i.e.

\begin{equation}
M_{\rm BH (XO-SMG)}= {{L_{\rm X (XO-SMG)}}\over{L_{\rm X (BL-SMG)}}} M_{\rm BH (BL-SMG)}
\end{equation}

Where $M_{\rm BH (XO-SMG)}$ and $L_{\rm X (XO-SMG)}$ are the black-hole
mass and X-ray luminosity of the X-ray obscured SMGs, and $M_{\rm BH
(BL-SMG)}$ and $L_{\rm X (BL-SMG)}$ are the black-hole mass and X-ray
luminosity of the broad-line SMGs.

This result initially suggests that there could be other factors beyond
the orientation of the obscuring torus that dictate the differences
between broad-line SMGs and X-ray obscured SMGs. Indeed, as can be seen
in Fig.~3, the X-ray luminosities for three of the broad-line SMGs are
$\approx$~10 times higher than the other broad-line SMGs and the X-ray
obscured SMGs. The average black-hole mass and Eddington ratio of these
X-ray luminous broad-line SMGs ($L_{\rm X}\approx10^{45}$~erg~s$^{-1}$)
is log($M_{\rm BH}/M_{\odot})$=~8.2--8.6 and $\eta\approx$~0.4--0.9,
with an overall average of log($M_{\rm BH}/M_{\odot})\approx$~8.4 and
$\eta\approx0.6$. These black-hole masses and Eddington ratios are
similar to the average found for SDSS quasars with
FWHM$<3500$~km~s$^{-1}$ [log($M_{\rm BH}/M_{\odot})$=~8.5;
$\eta\approx1.0$]; see Fig.~2. The X-ray luminous broad-line SMGs may
therefore be more closely related to the submm-detected quasars studied
by Page \etal (2001, 2004) and Stevens \etal (2005) and are potentially
more evolved objects than the X-ray obscured SMGs (e.g.,\ Coppin \etal
2008).

In contrast, the average black-hole mass and Eddington ratio of the
lower luminosity broad-line SMGs ($L_{\rm
X}\approx10^{44}$~erg~s$^{-1}$) is log($M_{\rm
BH}/M_{\odot})$=~7.8--8.2 and $\eta\approx$~0.1--0.3, depending upon
the geometry of the BLR gas (see \S3.2), with an overall average of
log($M_{\rm BH}/M_{\odot})\approx$~8.0 and $\eta\approx0.2$.  The X-ray
obscured SMGs lie in the same location of the $L_{\rm X}$--$L_{\rm
FIR}$ plane as the lower luminosity broad-line SMGs, providing evidence
that they are intrinsically similar systems where the nucleus is
obscured at optical wavelengths in the X-ray obscured SMGs (e.g.,\
Antonucci 1993). Under this scenario, the black-hole masses of the
X-ray obscured SMGs can be constrained from the Eddington ratios of the
broad-line SMGs; see Eqn.~1. Since all of the lower luminosity
broad-line SMGs have $\eta>0.1$, the estimated black-hole masses of the
X-ray obscured SMGs will be \hbox{log($M_{\rm
BH}$/$M_{\odot}$)~$<8.1$}, and the average black-hole mass will be
\hbox{log($M_{\rm BH}$/$M_{\odot}$)~$\approx$~7.8} for the average
Eddington ratio of $\eta\approx0.2$; we provide further evidence for
$\eta=0.2$ in \S4.

\subsection{The Black-hole--Galaxy mass relationship for X-ray obscured SMGs}

We can combine the black-hole mass constraints with host-galaxy masses
to extend the analyses of Borys \etal (2005) and explore the
black-hole--galaxy mass relationship for the X-ray obscured SMGs. The
stellar-mass constraints from Borys \etal (2005) were calculated for
the X-ray obscured SMGs in Alexander \etal (2005a,b) using rest-frame
UV--near-IR observations (ground-based, {\it Hubble Space Telescope},
and {\it Spitzer} imaging). Adopting appropriate mass-to-light ratios
for the stellar populations in these systems ($L_{\rm
K}/M\approx$~3.2), Borys \etal (2005) found a mean stellar mass of
$2.5\times10^{11}$~$M_{\odot}$ for all of the SMGs. However, excess
rest-frame UV or near-IR emission beyond that expected from starlight
indicates that AGN activity is probably contaminating the stellar
continuum in four galaxies (see Table~3 of Borys \etal
2005). Furthermore, the sample included three $z\simlt1$ galaxies,
which have lower redshifts than the broad-line SMGs. Therefore, when
calculating the average stellar masses here we have only considered the
six $z>1.8$ X-ray obscured SMGs that do not have a UV or near-IR excess
in Borys \etal (2005), giving a mean stellar mass of
$2.2\times10^{11}$~$M_{\odot}$, a factor $\approx$~2 less than the mean
of all of the $z>1.8$ X-ray obscured SMGs in Borys \etal (2005; $M_{\rm
GAL}\approx4.5\times10^{11}$~$M_{\odot}$). This stellar mass estimate
is consistent with the galactic masses estimated from the CO dynamics
[$(1.2\pm1.5)\times10^{11}$~$M_{\odot}$; Greve \etal 2005; see Swinbank
\etal 2004, 2006 for similar H$\alpha$ dynamical constraints]. The
Maraston (2005) stellar population models, which take into account
thermally pulsating asymptotic giant branch (TP-AGB) stars, would
predict similar stellar masses if the starbursts in SMGs are long
lived, as typically assumed. However, even in a less likely scenario
where the starbursts in SMGs are short lived (i.e.,\ $\simlt$~10~Myrs;
which leads to improbably high space densities for the $z\approx$~0
descendants of SMGs), the stellar masses estimated from the Maraston
(2005) model are still only $\simlt$~3 times lower than those given
here (see Swinbank \etal 2008 for a detailed discussion).

The local black-hole--galaxy mass relationship that we have adopted
here is based on the galaxy spheroid rather than the entire galaxy. The
CO dynamical constraints of Greve \etal (2005) trace the mass within
the central $\approx2$~kpc of SMGs, where the majority of the CO gas
lies (e.g.,\ Tacconi \etal 2006), while the stellar masses will include
the whole galaxy. Since the effective radius for early-type galaxies
with $M_{\rm GAL}>10^{11}$~$M_{\odot}$ is $>4$~kpc [see Fig.~2 of
Desroches \etal 2007 for $M_{\rm R}\simgt$~--21.5 ($\approx$$L_*$);
Bell \etal 2003], the CO gas is tracing scales smaller than the
spheroid and therefore provides a lower limit to the spheroid mass. We
also note here that the CO dynamical constraint includes all baryonic
material, not just the stars, however, the gas is likely to be
ultimately converted to stars in these systems. Similarly, the stellar
mass is $\approx$~2 times the CO dynamical mass and corresponds to the
entire galaxy. However, since the surface-brightness distributions of
most SMGs are reasonably well fitted with a spheroid-like $r^{1/4}$
profile (Borys \etal 2008), the stellar masses of these galaxies are
likely to be dominated by the spheroid. Furthermore, we note that if
SMGs are massive elliptical galaxies in formation (e.g.,\ Blain \etal
2004; Swinbank \etal 2006; Tacconi \etal 2006), as commonly assumed,
then any non-spheroid component in the host galaxy is likely to be
subsumed within the existing spheroid component. In this scenario, the
entire stellar mass would ultimately represent the final galaxy
spheroid mass of the system.

In Fig.~4 we plot the black-hole versus galaxy mass for the $z>1.8$
X-ray obscured SMGs, with the objects with excess emission removed; the
average estimated black-hole mass of these objects is the same as that
given for the full sample in \S3.4 [{log($M_{\rm
BH}$/$M_{\odot}$)~$\approx$~7.8}; i.e.,\ log($M_{\rm
BH,Edd}$/$M_{\odot}$)~$\approx7.1$]. The average black-hole--galaxy
mass ratio of the X-ray obscured SMGs is $\approx2.9\times10^{-4}$ and
$\approx5.3\times10^{-4}$ for the average stellar mass and CO dynamical
mass, respectively. Assuming the black hole--spheroid mass relationship
of H{\"a}ring \& Rix (2004), the SMGs lie $\approx$~3--5 times below
that found for comparably massive normal galaxies in the local
Universe, with considerable uncertainty, depending upon whether the
average CO dynamical mass or stellar mass is used. If the Marconi \&
Hunt (2003) relationship adopted by Borys \etal (2005) is used then the
SMGs would lie $\approx$~1.4 times further below the local
relationship. The average black-hole--galaxy mass ratio that we
estimate for the X-ray obscured SMGs is an order of magnitude higher
than that estimated in Borys \etal (2005). The primary reason for this
difference is due to using $\eta=0.2$ (rather than $\eta=1.0$ assumed
by Borys \etal 2005), although the exclusion of the SMGs with excess
emission also contributes to the higher value determined here.

%
%%%%%%%%%%%%%%%%%%%%%%%%%%%%%%%%%%%%%%%%%%%%%%%%%%%%%%%%%%%%%%%%%%%%%%
\section{Comparison to nearby ULIRGs}
%%%%%%%%%%%%%%%%%%%%%%%%%%%%%%%%%%%%%%%%%%%%%%%%%%%%%%%%%%%%%%%%%%%%%%
%

Our analyses in \S3 provided the first quantitative estimates of the
black-hole masses and black-hole--galaxy mass relationship of the SMG
population. We estimated the black-hole masses of the X-ray obscured
SMGs under the assumption that they are the obscured counterparts of
the lower-luminosity broad-line SMGs, finding \hbox{log($M_{\rm
BH}$/$M_{\odot}$)~$\approx$~7.8} for $\eta\approx$~0.2.  However, there
are a number of uncertainities that could be better constrained. First,
it would be useful to have independent black-hole mass constraints for
the X-ray obscured SMGs, to reduce the need to make assumptions about
their relationship with the broad-line AGNs and to minimise the
potential inclination-dependent effects on the widths of the broad
emission lines. Second, the determination of the black-hole masses of
the X-ray obscured SMGs depend upon accurate intrinsic X-ray
luminosities, which relies on correcting the observed X-ray
luminosities for the effect of absorption. We deal with these two
issues here by studying nearby ULIRGs that host obscured AGNs. The
advantage of this approach is that the observations of nearby ULIRGs
are often more extensive and of a considerably higher signal-to-noise
ratio than that obtained for distant SMGs, although clearly the
disadvantage is that we will be studying objects at a different epoch
to the SMGs. However, the similarity in the properties of SMGs and the
nearby ULIRGs, and the fact that they both appear to be major-merger
induced events suggests that SMGs could be ``scaled-up'' versions of
ULIRGs (e.g.,\ Tacconi \etal 2006), providing some support for our
general approach (see \S2.2).

\subsection{Are the Eddington ratios smaller for the X-ray obscured SMGs?}

Although the presence of obscuration prevents the opportunity to
measure the black-hole properties of the X-ray obscured SMGs, the
identification of broad Pa$\alpha$ emission provides the chance to
constrain $M_{\rm BH}$ and $\eta$ in a small number of nearby ULIRGs
hosting obscured AGNs (e.g.,\ Veilleux \etal 1997, 1999). The
broad-line emission from these sources is thought to be observed
through the atmosphere of the dusty torus, which is assumed to be
optically thin at near-IR wavelengths. A benefit of focusing on
obscured AGNs is that any potential inclination-dependent affects on
the black-hole mass and Eddington ratio will be minimised since the BLR
is unlikely to be seen close to pole on.

Out of the sample of 64 nearby obscured ULIRGs observed by Veilleux
\etal (1997, 1999), six have broad Pa$\alpha$ emission; see Table~2 for
their properties. We determined the black-hole masses of these objects
using the Greene \& Ho (2005) virial black-hole mass estimator by
correcting the Pa$\alpha$ luminosity to that expected for H$\alpha$
using the Case B approximation (H$\alpha$/Pa$\alpha$~=~8.6; e.g.,\
Hummer \& Storey 1987); see Fig.~1. In Fig.~2 we show the mass
accretion rate versus virial black-hole mass for the four obscured
ULIRGs that have sensitive X-ray constraints. The mass accretion rates
were estimated from the absorption-corrected X-ray luminosity following
\S3.3, and we converted to the 0.5--8.0~keV band assuming $\Gamma=1.8$,
when necessary. The Eddington ratios of these obscured ULIRGs range
from $\eta=$~0.1--0.5, with an average of $\eta=0.2^{+0.2}_{-0.1}$. The
average Eddington ratio is similar to that found for the
lower-luminosity broad-line SMGs, further suggesting that
$\eta\approx0.2$ is a good choice for the X-ray obscured SMGs.

%
%%%%%%%%%%%%%%%%%%%%%%%%%%%%%%%%%%%%%%%%%%%%%%%%%%%%%%%%%%%%%%%%%%%%%%
% TABLE 2
%%%%%%%%%%%%%%%%%%%%%%%%%%%%%%%%%%%%%%%%%%%%%%%%%%%%%%%%%%%%%%%%%%%%%%
%

\begin{table}[!t]
 \centering
  \caption{Obscured ULIRGs with Broad Pa$\alpha$ Emission}
  \begin{tabular}{@{}lccccccccc@{}}
  \hline
             & $d_L$  &   log($L_{\rm Pa\alpha}$)$^{\rm a}$ &
             FWHM$_{Pa\alpha}$ & log($M_{\rm BH}$) & log($L_{\rm
             X}$)$^{\rm a,b}$ & $M_K$$^{\rm a}$ & \\
 Name        & (Mpc) &   (erg~s$^{-1}$) & (km~$s^{-1}$)    &             ($M_{\odot}$)     & (erg~s$^{-1}$)  & (mag) & Refs\\
 \hline
I05189-2524  & 169  &     41.8  &    $2600$  &    7.4 & 43.3    & -24.7 & 1,2,3\\
I13305-1739  & 695  &     41.9  &    $2900$  &    7.6 & $\dots$ & -26.0 &1,3\\
PKS1345+12   & 563  &     42.3  &    $2600$  &    7.7 & 43.4    & -26.1 & 3,4,5\\
I20460+1925  & 868  &     43.1  &    $2900$  &    8.2 & 44.2    & $\dots$ & 4,6\\
I23060+0505  & 831  &     42.9  &    $2000$  &    7.8 & 44.3    & -25.5 & 3,4,7\\
I23498+2423  & 1036 &     43.0  &    $3000$  &    8.2 & $\dots$ & -26.9 & 3,4\\
\hline
\vspace{0.01in}
\end{tabular}
\begin{minipage}{85mm}
$^a$~Re-calculated from the published value using the given luminosity distance.\\
$^b$~Rest-frame 2--10~keV band, corrected for absorption.\\
{\sc References:} --
(1) Veilleux \etal (1999); (2) Severgnini \etal (2001); (3)
Veilleux \etal (2002); (4) Veilleux \etal (1997); (5) O'Dea \etal (2000); (6)
Ogasaka \etal (1997); (7) Brandt \etal (1997).\\
\end{minipage}
\end{table}

%
%%%%%%%%%%%%%%%%%%%%%%%%%%%%%%%%%%%%%%%%%%%%%%%%%%%%%%%%%%%%%%%%%%%%%%
% TABLE 3
%%%%%%%%%%%%%%%%%%%%%%%%%%%%%%%%%%%%%%%%%%%%%%%%%%%%%%%%%%%%%%%%%%%%%%
%

\begin{table}[!t]
 \centering
  \caption{$d<200$~Mpc ULIRGs Hosting AGN Activity}
  \begin{tabular}{@{}lccccccccc@{}}
  \hline
             & $d_L$$^{\rm a}$  &   log($L_{\rm FIR}$)$^{\rm a}$ &
             log($L_{\rm X}$)$^{\rm b,c}$ & log($N_H$) & \\
 Name        & (Mpc) &  ($L_{\odot}$) & (erg~s$^{-1}$) & (cm$^{-2}$)    & Refs\\
 \hline
I05189-2524 &169 &12.0 &  43.5 & 22.8& 1 \\
UGC5101     &159 &11.9 &  42.8 & 24.1& 2 \\
MKN231      &172 &12.3 &  44.2 & 24.3& 3 \\
MKN273      &155 &12.1 &  43.0 & 23.8& 4 \\
I20551-4250 &173 &11.9 &  43.1 & 23.9& 5 \\
I23128-5919 &178 &12.0 &  42.4 & 22.8& 5 \\
\hline
\vspace{0.01in}
\end{tabular}
\begin{minipage}{85mm}
$^a$~Taken from Sanders \etal (2003).\\
$^b$~Re-calculated from the published value using the given luminosity distance.\\
$^c$~Rest-frame 2--10~keV band, corrected for absorption.\\
{\sc References:} --
(1) Severgnini \etal (2001); (2) Imanishi \etal (2003); (3) Braito
\etal (2004); (4) Balestra \etal (2005); (5) Franceshini \etal (2003).\\
\end{minipage}
\end{table}

\subsection{Have the intrinsic luminosities of the X-ray obscured SMGs
  been underestimated?}

The estimated black-hole masses of the X-ray obscured SMGs are heavily
dependent on how accurately the intrinsic luminosity of the AGN (and
hence the Eddington-limited black-hole mass; Alexander \etal 2005a) is
calculated. The X-ray spectra produced for the X-ray obscured SMGs in
Alexander \etal (2005b) showed that the majority of the AGNs are
heavily obscured ($\approx$~80\% have $N_{\rm H}>10^{23}$~cm$^{-2}$),
indicating that potentially large absorption corrections may need to be
applied to calculate the intrinsic X-ray luminosities. Given the poor
photon statistics for the X-ray spectra of individual objects it is
useful to verify that the adopted absorption corrections used in
Alexander \etal (2005b) are typical for obscured AGNs. This is
particularly important for the most heavily obscured AGNs as the
corrections become increasingly model dependent as the absorption
approaches and exceeds Compton-thick obscuration (i.e.,\ $N_{\rm
H}=1.5\times10^{24}$~cm$^{-2}$) due to uncertain contributions from
reflected and scattered AGN components (e.g.,\ Mushotzky \etal 1993;
Matt \etal 2000).

In Fig.~5 we show the applied absorption corrections for the $z>1.8$
X-ray obscured SMGs in Alexander \etal (2005b) and compare them to
those adopted in other X-ray spectral analysis studies of low and high
redshift ULIRGs. The absorption corrections applied to the SMGs are in
good agreement with those used in other studies, indicating that unless
the X-ray obscured SMGs are more heavily obscured than estimated in
Alexander \etal (2005b), their intrinsic X-ray luminosities are not
significantly underestimated (i.e.,\ $L_{\rm
X}\approx10^{44}$~erg~s$^{-1}$). Faint {\it Spitzer}-IRS spectroscopy
also shows that the AGN activity in SMGs is typically weak at mid-IR
wavelengths, in good agreement with that implied by their X-ray
luminosities, assuming typical X-ray--mid-IR conversions for AGNs
(e.g.,\ Men{\'e}ndez-Delmestre \etal 2007; Valiante \etal 2007; Pope
\etal 2008).

%
% Figure 5: absorption corrections
%
\begin{figure}[!t]
\plotone{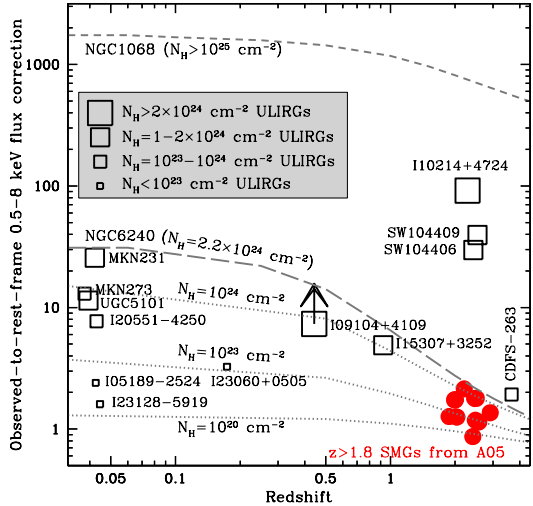}
\caption{X-ray flux absorption corrections versus redshift for X-ray
obscured AGNs in ULIRGs at low and high redshift (taken from Brandt
\etal 1997; Franceschini \etal 2000, 2003; Severgnini \etal 2001;
Imanishi \etal 2003; Ptak \etal 2003; Braito \etal 2004; Alexander
\etal 2005c; Balestra \etal 2005; Mainieri \etal 2005; Polletta \etal
2006) and the $z>1.8$ X-ray obscured SMGs studied in Alexander \etal
(2005b).  The corrections that would need to be made for two nearby
Compton-thick AGN NGC~1068 (dashed curve; using Matt \etal 1997 model)
and NGC~6240 (long dashed curve; using Vignati \etal 1999 model) are
shown in addition to the models adopted by Alexander \etal (2005b) for
lower amounts of obscuration (dotted curves); the assumed column
densities are indicated. The X-ray derived column densities for the
ULIRGs are indicated using different symbol sizes. The negative
$K$-corrections for AGNs in the X-ray band mean that the absorption
corrections are not large for high-redshift objects unless they are
heavily Compton thick (i.e.,\ $N_{\rm H}\gg10^{24}$~cm$^{-2}$). The
good agreement between the absorption corrections used in Alexander
\etal (2005b) and those of obscured ULIRGs in the literature indicate
that the intrinsic X-ray luminosities of the X-ray obscured SMGs are
unlikely to be significantly underestimated.}
\end{figure}

%
% Figure 6: Comparison to d<200 Mpc ULIRGs
%
\begin{figure}[!t]
%\vspace{0.1in}
\plotone{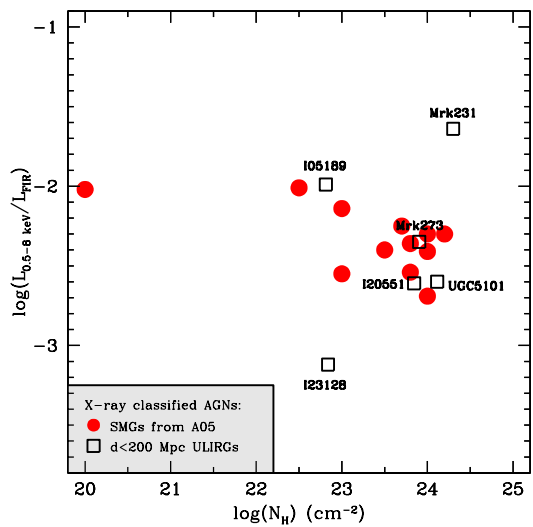}
\caption{Unabsorbed 0.5--8.0~keV-to-far-IR luminosity ratio versus
X-ray derived column densities for the $z>1.8$ X-ray obscured SMGs from
Alexander \etal (2005b) and the six $d<200$~Mpc ULIRGs hosting obscured
AGN activity (see \S4.2). The far-IR luminosities for the X-ray
obscured SMGs are from Alexander \etal (2005b) and are calculated using
the radio-to-far-IR relationship, and the far-IR luminosities for the
$d<200$~Mpc ULIRGs are taken from Sanders \etal (2003). The column
density and X-ray-to-far-IR luminosity ratio distributions of both
populations are similar.}
\end{figure}

The poor photon statistics for the individual X-ray spectra of the
X-ray obscured SMGs in Alexander \etal (2005b) can also lead to
significant uncertainties on the measured absorption. In an attempt to
minimise this, Alexander \etal (2005a) constructed composite rest-frame
2--20~keV spectra for the SMGs grouped into three different absorption
bands: $N_{\rm H}<10^{23}$~cm$^{-2}$, $N_{\rm
H}=$~(1--5)~$\times10^{23}$~cm$^{-2}$, and $N_{\rm
H}>5\times10^{23}$~cm$^{-2}$. These composite X-ray spectra validated
the individual X-ray spectral analyses and showed that the average
intrinsic X-ray luminosity of the most heavily obscured AGNs is
comparable to that of the less obscured AGNs (where the absorption
corrections are smaller), as expected if all sources have the same
intrinsic AGN properties but some are more heavily obscured (e.g.,\
Antonucci 1993; Mushotzky \etal 1993; Smith \& Done 1996). However, the
composite X-ray spectra still only had moderately good photon
statistics ($\approx$~600--1000 X-ray counts; see Fig.~7 of Alexander
\etal 2005b), and validation of the estimated column densities would be
useful. Better statistics could be achieved by the identification of
more SMGs hosting AGN activity in the {\it Chandra} Deep Fields, which
should be possible with the advent of the ultra-deep
450~$\mu$m/850~$\mu$m SCUBA2 Cosmology Legacy Surveys, or from deeper
X-ray exposures in these fields.\footnote{See
http://www.jach.hawaii.edu/JCMT/surveys/Cosmology.html for details of
the SCUBA2 Cosmology Legacy Surveys.}  However, in the immediate
absence of these two possibilities, a reasonable alternative is to
compare the properties of the AGNs in the X-ray obscured SMGs to the
AGNs in nearby ULIRGs, where the signal-to-noise ratio of the data is
higher.

The initial ULIRG sample selected for this analysis comprises the ten
ULIRGs at $d<200$~Mpc from the {\it IRAS} Revised Bright Galaxy Sample
(RBGS; Sanders \etal 2003). By restricting our comparison to the
closest ULIRGs we ensure homogenous, high signal-to-noise hard X-ray
data (all of the galaxies have {\it Chandra} data and nine galaxies
have {\it XMM-Newton} data); the X-ray coverage of ULIRGs at
$d>200$~Mpc is considerably poorer. Due to the larger effective area of
{\it XMM-Newton} over {\it Chandra} at $>6$~keV, where the Fe~K$\alpha$
emission line and photo-electric cutoff for heavily obscured nearby AGN
are present, the X-ray properties of the AGNs used in these analyses
are only taken from {\it XMM-Newton} observations (Severgnini \etal
2001; Franceschini \etal 2003; Imanishi \etal 2003; Braito \etal 2004;
Balestra \etal 2005).

Six ($60^{+35}_{-24}$\%; errors calculated from Gehrels 1986) of the 10
$d<200$~Mpc ULIRGs show evidence for AGN activity at X-ray energies and
all are heavily obscured ($N_{\rm H}>10^{22}$~cm$^{-2}$); see Table~3
for the properties of these objects. In Fig.~6 we show the
absorption-corrected X-ray-to-far-IR luminosity ratio versus the X-ray
measured absorption column density of these sources and compare them to
the $z>1.8$ X-ray obscured SMGs from Alexander \etal (2005b). Although
selection effects may be present, the AGN properties of these two
populations are similar [average log($N_{\rm H}$)~$\approx$~23.8
  cm~$^{-2}$ and log($L_{\rm X}/L_{\rm FIR}$)~$\approx$~--2.5],
indicating that the properties of the AGNs in the SMGs are comparable
to those found in the nearby ULIRGs, despite being significantly more
distant and luminous.

Lastly, we briefly consider the possibility that a significant fraction
of the AGN luminosity is undetected in the $<10$~keV observed-frame
data (e.g.,\ due to the presence of extreme Compton-thick absorption;
$N_{\rm H}>10^{25}$~cm$^{-2}$). Current ultra-hard
($\approx$~10--100~keV) X-ray observatories are not sensitive enough to
place stringent constraints on the $>10$~keV emission from X-ray
obscured SMGs. However, we note that the six X-ray detected AGNs in the
$d<200$~Mpc ULIRG sample have $>10$~keV constraints from {\it
BeppoSAX}-PDS and the {\it Swift}-BAT survey (e.g.,\ Severgnini \etal
2001; Braito \etal 2004; Markwardt \etal 2005; Dadina \etal 2007) which
already rule out the presence of an AGN component $>5$ times more
luminous than that estimated from the {\it XMM-Newton} data for these
sources; see Fig.~3.

The similarity in the column densities and X-ray-to-far-IR luminosity
ratios of the $d<200$~Mpc ULIRGs and the X-ray obscured SMGs suggest
that the intrinsic X-ray luminosities of the SMGs are unlikely to be
significantly higher than those estimated in Alexander \etal
(2005b). We therefore conclude that the average X-ray derived
black-hole masses of the X-ray obscured SMGs are unlikely to be
significantly higher (i.e.,\ by factors $\simgt2$) than the constraints
given in \S3.

\subsection{The Black-hole--Galaxy mass relationship for nearby obscured ULIRGs}

With black-hole masses for the obscured ULIRGs, we can take the same
approach as for the SMGs and determine the black-hole--galaxy mass
relationship for these galaxies. In Fig.~4 we show the black-hole
versus galaxy mass for the obscured ULIRGs with broad Pa$\alpha$
emission; the statistics are too poor for the $d<200$~Mpc ULIRG sample
since only three objects have stellar-mass constraints and only one
object has a black-hole mass constraint. The galaxy masses were
calculated using AGN-subtracted $K$-band host-galaxy magnitudes from
Veilleux \etal (2002), adopting the same mass-to-light ratio as that
used by Borys \etal (2005; $L_K/M=3.2$) for the SMGs; see Table~2.  The
average black-hole-to-galaxy mass ratio of the obscured ULIRG sample is
consistent with those of the SMGs. This is significant since both the
black hole and galaxy mass constraints were determined directly for
these ULIRGs, reducing the number of potential uncertainties in the
derived $M_{\rm BH}$--$M_{\rm GAL}$ ratio.  As for the SMGs, the
stellar masses of the ULIRGs are for the entire galaxy, however, since
$\approx$~73\% of ULIRGs have an elliptical-like $r^{1/4}$
surface-brightness profiles (Veilleux \etal 2002), the stellar mass
will be dominated by the galaxy spheroid. Furthermore, the luminosity,
$R$-band axial ratio, velocity dispersion distribution and location of
nearby ULIRGs on the fundamental place already closely resemble those
of intermediate-mass elliptical galaxies (e.g.,\ Genzel \etal 2001;
Veilleux \etal 2002; Dasyra \etal 2006), suggesting that the stellar
mass is tracing the ultimate mass of the system.

We therefore conclude that we see qualitatively similar behavior in
both the SMGs and the nearby ULIRGs. Since the nearby ULIRGs are easier
to study in detail than the more distant SMGs, they could potentially
provide future insight into the evolutionary status of these galaxy
populations (e.g.,\ Hao \etal 2005; Kawakatu \etal 2006).

%
%%%%%%%%%%%%%%%%%%%%%%%%%%%%%%%%%%%%%%%%%%%%%%%%%%%%%%%%%%%%%%%%%%%%%%
\section{Discussion}
%%%%%%%%%%%%%%%%%%%%%%%%%%%%%%%%%%%%%%%%%%%%%%%%%%%%%%%%%%%%%%%%%%%%%%
%

%
% Figure 7: Comparison with other source populations
%
\begin{figure*}[!t]
\centerline{\includegraphics[angle=0,width=12cm]{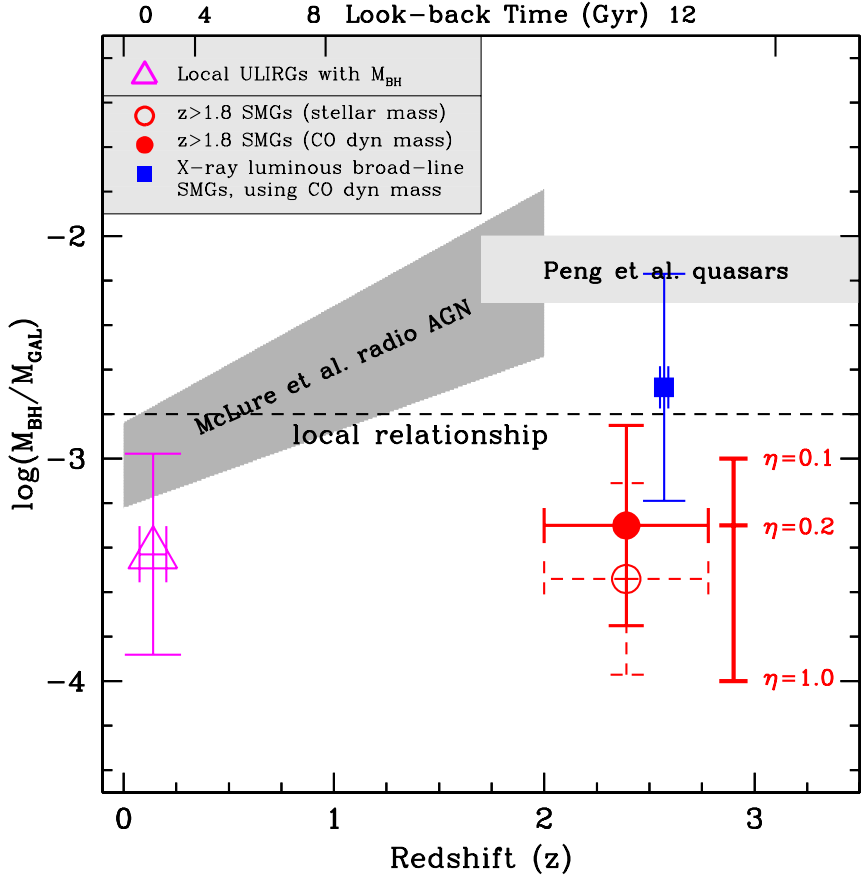}}
\figcaption{Black-hole--host-galaxy mass ratio versus redshift for the
$z>1.8$ X-ray obscured SMGs, X-ray luminous broad-line SMGs, and the
obscured ULIRGs with broad Pa$\alpha$ emission. The $z>1.8$ X-ray
obscured SMGs are plotted using the stellar mass and the CO dynamical
mass constraints. The dashed line indicates the local black
hole--spheroid mass relationship from H{\"a}ring \& Rix (2004), the
dark-shaded region shows the radio galaxy constraints from McLure \etal
(2006), and the light-shaded region indicates the constraints for
$z\approx$~2 quasars from Peng \etal (2006). The error bars correspond
to $1\sigma$ uncertainties. The X-ray obscured SMGs lie marginally
below the local relationship (factor $\approx$~3--5), as also found for
the nearby obscured ULIRGs, and they lie a factor $\simgt$~10 below
that found for $z\approx$~2 radio galaxies and quasar population. This
suggests that SMGs are identified at an earlier stage in the evolution
of quasars and massive galaxies. The X-ray luminous broad-line SMGs are
statistically consistent with the local relationship, assuming the CO
dynamical mass constraints from the SMGs and the average Eddington
ratio of $\eta=0.6$. We discuss the overall constraints in \S5.}
\vspace{0.15in}
\end{figure*}

The primary aim of this paper has been to provide quantitiative
constraints on the black-hole masses and evolutionary status of the SMG
population. Using virial black-hole mass estimates, we constrain the
black-hole masses of $z>1.8$ X-ray obscured SMGs to be log($M_{\rm
BH}/M_{\odot}$)~$\approx$~7.8 for $\eta=0.2$.  When combined with
host-galaxy mass constraints, the black holes of the X-ray obscured
SMGs appear to be $\approx$~3--5 times smaller than those found for
comparably massive galaxies in the local Universe, implying that the
growth of the black hole lags that of the host galaxy in SMGs. We
provided corroborating evidence for this result from analyses of nearby
obscured ULIRGs, potentially the $z\approx$~0 analogs of the X-ray
obscured SMGs.  In this final section we compare our results to those
found for $z\approx$~2 radio galaxies and quasars and explore the
evolutionary status and black-hole growth of the X-ray obscured SMGs.

\subsection{Comparison to $z\approx$~2 optically luminous Quasars and
  Radio galaxies}

Our results for the X-ray obscured SMGs are statistically inconsistent
with investigations of $z\approx$~2 radio galaxies and luminous
quasars, where it has been found that the black holes hosted by these
objects are up-to $\approx$~6 times larger than those found for
comparably massive galaxies in the local Universe (e.g.,\ McLure \etal
2006; Peng \etal 2006).  We demonstrate this in Fig.~7, where we show
the black-hole--galaxy mass ratio versus redshift for the X-ray
obscured SMGs and compare them to the $z\approx$~2 radio galaxies and
quasars and the $z\approx$~0 obscured ULIRGs with black-hole mass
constraints. To be consistent with the results of Peng \etal (2006),
the black holes in the SMGs would need to be $\simgt$~10 times more
massive than those estimated here (i.e.,\ $M_{\rm
BH}\simgt10^9$~$M_{\odot}$). However, none of the broad-line SMGs in
our sample have such high black-hole masses, even allowing for
potential BLR inclination angle effects.  Furthermore, black holes with
$M_{\rm BH}\simgt10^9$~$M_{\odot}$ are rare even in the local Universe,
with a space density of $\phi\approx10^{-5}$~Mpc$^{-3}$ (e.g.,\ McLure
\& Dunlop 2004), which is comparable to the {\rm observed} space
density of $z\approx$~2 SMGs and about an order of magnitude lower than
the SMG space density {\it corrected} for a 300~Myr submm-bright phase
($\phi\approx1.3\times10^{-4}$~Mpc$^{-3}$; see \S4.2 of Swinbank \etal
2006). These analyses show that the $z\approx$~0 descendants of typical
SMGs cannot host massive black holes of $M_{\rm
BH}\simgt10^9$~$M_{\odot}$ and therefore suggests that the most
luminous radio galaxies and quasars investigated by McLure \etal (2006)
and Peng \etal (2006) have a different evolutionary path to the
SMGs. Indeed, given the significantly larger fraction of radio galaxies
that are submm-detected at $z>2.5$ when compared to the submm-detected
fraction of radio galaxies at $z<2.5$ ($\simgt$~75\% at $z>2.5$ versus
$\approx$~15\% at $z<2.5$; Archibald \etal 2001), it seems likely that
these galaxies (which represent the most massive galaxies at every
epoch; Seymour \etal 2007) underwent their major star-formation phases
at higher redshifts than the SMG population. We note that this
conclusion is also consistent with detailed spectral fitting of the
stellar populations of nearby massive galaxies (e.g.,\ Nelan \etal
2005; Panter \etal 2007).

However, it is also possible that the variations in the $M_{\rm
BH}$--$M_{\rm GAL}$ relationship at $z\approx$~2 are due to selection
effects, as first suggested by Lauer \etal (2007). The selection of the
most luminous quasars and radio galaxies will clearly identify the most
massive black holes since the Eddington limit restricts the luminosity
of small accreting black holes, potentially causing a bias toward
objects with large $M_{\rm BH}$--$M_{\rm GAL}$ ratios.  It is not clear
that the selection of SMGs will cause an opposite bias since there is
not an equivalent stellar-mass based Eddington-limited restriction in
star-forming galaxies. However, empirically, the galaxies that are
undergoing the most intense star formation at $z\approx$~2 also appear
to be massive galaxies at this epoch (e.g.,\ Papovich \etal 2006),
suggesting that the selection of SMGs will cause a bias toward the
identification of massive galaxies. However, this selection is probably
not as closely coupled to a bias in the $M_{\rm BH}$--$M_{\rm GAL}$ as
the selection of radio galaxies and quasars.

\subsection{The Evolutionary Status and Black-Hole growth of SMGs}

Our results show that the $z\approx$~0 descedants of SMGs cannot host
black holes with $M_{\rm BH}\simgt10^9$~$M_{\odot}$. Indeed, on the
basis of dynamical mass estimates, stellar luminosities, and
star-formation lifetimes, Swinbank \etal (2006) have shown that the
local descendants of SMGs are likely to be $\simgt3L_{\rm *}$
early-type galaxies [$M_{\rm GAL}\simgt3\times10^{11}$~$M_{\odot}$,
assuming the Bell \etal (2003) luminosity and stellar-mass functions],
with space densities of $\approx10^{-4}$~Mpc$^{-3}$.  We therefore
expect the $z\approx$~0 descendants of SMGs to host black holes with
$M_{\rm BH}\simgt4\times10^{8}$~$M_{\odot}$, comparable to those of the
X-ray luminous broad-line SMGs. This would suggest that the black holes
of SMGs and their descendants typically only need to grow by a factor
of $\approx$~6 by the present day.  Can we account for this black-hole
growth in the estimated submm-bright lifetime of SMGs
($\approx$~100--300~Myrs)?

Assuming that the black holes in the X-ray obscured SMGs are growing at
$\eta\approx$~0.2 (the average Eddington rate estimated in \S3.4), it
will take $\approx$~400~Myrs for them to increase in mass by a factor
$\approx$~6. Since we are likely to be observing a typical SMG halfway
through its submm-bright phase, this lifetime is longer than the
remaining gas consumption timescale (i.e.,\ $\approx$~150~Myrs, taking
the $\approx$~300~Myr total lifetime estimated by Swinbank \etal
2006). However, as mentioned in \S3.4, the Eddington ratios derived for
the broad-line SMGs should only be considered indicative and it is
possible that the X-ray obscured SMGs are growing their black holes at
$\eta>0.2$. For example, taking the latest derivation of bolometric
corrections from Vasudevan \& Fabian (2007) for high accretion rate
AGNs, the estimated Eddington ratio would be $\eta\approx$~0.4 (i.e.,\
$\kappa_{2-10 keV}\approx$~70; see \S3.3), which is sufficient for a
black hole to grow by a factor $\approx$~6 in $\approx$~200~Myrs and
broadly consistent with current estimates for the submm-bright lifetime
of SMGs.

We can also take the alternative approach adopted by Kawakatu \etal
(2007b) for local ULIRGs, and assume that the bolometric luminosity
produced by SMGs is entirely due to AGN activity.  Under this
hypothesis we would estimate Eddington ratios of $\eta\approx$~2
(i.e.,\ for a mean $L_{\rm FIR}\approx5\times10^{12}$~$M_{\odot}$; see
Fig.~3). However, we consider this scenario too extreme since all of
the current evidence indicates that star-formation dominates the
bolometric luminosity of typical SMGs, as witnessed by (1) the
correlation between far-IR luminosity and reddening corrected H$\alpha$
SFR (e.g.,\ Swinbank \etal 2004; Takata \etal 2006), (2) the strong PAH
features seen in their mid-IR spectra and the correlation between the
luminosity of the PAH emission and the total IR luminosity (e.g.,\
Men{\'e}ndez-Delmestre \etal 2007; Pope \etal 2008), (3) the extended
radio emission in high resolution ($<4$~kpc) maps (e.g.,\ Chapman \etal
2004; Biggs \& Ivison 2008), and (4) the absorption-corrected X-ray to
far-IR luminosities, which indicate that AGN activity contributes
$\approx$~10\% of the bolometric luminosity (e.g.,\ Alexander \etal
2005b). All of these facts suggest that the bolometric luminosity of
SMGs is dominated by star-formation, and therefore this determination
of the Eddington ratio provides a firm upper limit.

Finally, we consider the properties of the X-ray luminous ($L_{\rm
  X}\approx10^{45}$~erg~s$^{-1}$) broad-line SMGs; see \S3.4.  These
  objects have more massive black holes than the X-ray obscured SMGs
  and they might represent transition objects between typical SMGs and
  typical unobscured quasars (e.g.,\ Coppin \etal 2008); see Fig.~7 for
  constraints on their $M_{\rm BH}$--$M_{\rm GAL}$ ratio, assuming that
  they have host-galaxy masses comparable to the X-ray obscured
  SMGs. The location of the X-ray luminous broad-line SMGs in the
  $L_{\rm X}$--$L_{\rm FIR}$ plane indicates that these objects might
  be similar to the submm-detected quasars identified by Page \etal
  (2001, 2004) and Stevens \etal (2005).  It has been argued that
  submm-detected quasars represent a late stage in the evolution of
  SMGs (e.g.,\ Coppin \etal 2008), where accretion-related outflows
  from the black hole have started to remove gas and dust from the
  nucleus and the host galaxy (as suggested by most models; e.g.,\
  Sanders \etal 1988; Di Matteo \etal 2005; King 2005; Granato \etal
  2006; Hopkins \etal 2006a; Chakrabarti \etal 2007). A natural
  consequence of these accretion-related outflows should be a decrease
  in the fraction of luminous obscured AGNs. It is therefore
  interesting that current submm surveys have not yet identified a
  large population of luminous X-ray obscured quasars (see Bautz \etal
  2000, Mainieri \etal 2005, and Pope \etal 2008 for some objects),
  indicating that they are rare. The combination of the wide-area
  SCUBA2 Cosmology Legacy Survey (see Footnote 4) with moderately deep
  X-ray and IR coverage will provide the most definitive constraints on
  the ubiquity of submm-detected X-ray obscured quasars and test
  whether they are transition objects between typical SMGs and typical
  unobscured quasars.

\vspace{0.5in}
%
%%%%%%%%%%%%%%%%%%%%%%%%%%%%%%%%%%%%%%%%%%%%%%%%%%%%%%%%%%%%%%%%%%%%%%
\section{Conclusions}
%%%%%%%%%%%%%%%%%%%%%%%%%%%%%%%%%%%%%%%%%%%%%%%%%%%%%%%%%%%%%%%%%%%%%%
%

We have placed direct observational constraints on the black-hole
masses ($M_{\rm BH}$) and Eddington ratios ($\eta=L_{\rm bol}/L_{\rm
Edd}$) of the cosmologically important $z\approx$~2 SMG population, and
used measured host-galaxy masses to explore their evolutionary
status. Our main findings are the following:

\begin{enumerate}

\item We used the virial black-hole mass estimator to ``weigh'' the
  black holes in six SMGs with broad H$\alpha$ or H$\beta$ emission and
  found that they typically host black holes with log($M_{\rm
  BH}/M_{\odot}$)~$\approx$~8.0--8.4 and $\eta\approx$~0.2--0.5,
  depending on the geometry of the broad-line region gas. These black
  holes are $\approx$~0.4~dex smaller than those found for $z\approx$~2
  optically bright quasars in the SDSS with comparably ``narrow'' broad
  lines. See \S3.1--3.3.

\item We found that the lower-luminosity broad-line SMGs ($L_{\rm
  X}\approx10^{44}$~erg~s$^{-1}$) lie in the same location of the
  $L_{\rm X}$--$L_{\rm FIR}$ plane as more typical SMGs hosting X-ray
  obscured AGNs (X-ray obscured SMGs). These two subsets of the SMG
  population may be intrinsically similar, where the rest-frame optical
  nucleus is visible in the broad-line SMGs (as suggested by the
  unified AGN model). Under this hypothesis, the X-ray obscured SMGs
  host black holes of log($M_{\rm BH}/M_{\odot}$)~$\approx$~7.8 with
  $\eta=0.2$; we find corroborating evidence for $\eta\approx0.2$ from
  detailed analyses of nearby obscured ULIRGs with broad Pa$\alpha$
  emission, potentially the local analogs of the X-ray obscured SMGs.
  By comparison, the black holes and Eddington ratios of the X-ray
  luminous broad-line SMGs ($L_{\rm X}\approx10^{45}$~erg~s$^{-1}$) are
  log($M_{\rm BH}/M_{\odot}$)~$\approx$~8.4 and $\eta\approx$~0.6,
  suggesting that they are more evolved objects than typical SMGs. See
  \S3.4--3.5, \& \S4.1.

\item We demonstrated that the X-ray absorption corrections and
  estimated absorbing column densities of the X-ray obscured SMGs are
  consistent with those typically found for nearby ULIRGs. This
  suggests that we have not significantly underestimated the intrinsic
  X-ray luminosities (and therefore the black-hole masses) of the X-ray
  obscured SMGs. See \S4.2.

\item We combined the black-hole mass constraints with measured
  host-galaxy masses [$M_{\rm
  GAL}\approx$~(1--2)~$\times10^{11}$~$M_{\odot}$]. We found that X-ray
  obscured SMGs typically host black holes $\simgt$~3 times smaller
  than those found in comparably massive normal galaxies in the local
  Universe, albeit with considerable uncertainty, and $\simgt$~10 times
  smaller than those predicted for $z\approx$~2 populations. These
  results imply that the growth of the black hole lags that of the host
  galaxy in SMGs, in stark contrast with that previously found for
  optical and radio selected luminous quasars at $z\approx$~2, which is
  probably due to the SMGs being selected at an earlier evolutionary
  stage than the quasars.  We also found that the growth of the black
  hole lags that of the host galaxy in local ULIRGs. See \S3.5, \S4.3,
  \& \S5.

\item We provided evidence that the black holes of the descendants of
  SMGs are likely to be $M_{\rm BH}\approx4\times10^{8}$~$M_{\odot}$,
  rather than the $M_{\rm BH}\simgt10^{9}$~$M_{\odot}$ that would be
  required if they were to lie on the $M_{\rm BH}$--$M_{\rm GAL}$
  relationship found for $z\approx$~2 radio galaxies and quasars. This
  implies that SMGs only need to grow their black holes by a factor
  $\approx$~6 by the present day, which can be achieved within current
  estimates for the submm-bright lifetime of SMGs, provided that the
  black holes are growing at rates close to the Eddington limit. See
  \S5.

\end{enumerate}

The up-coming wide and deep SCUBA2 Cosmology Legacy Surveys will
provide larger samples of SMGs with which to test these results and
provide further insight into the relationship between the growth of
massive black holes and their host galaxies in this cosmologically
important $z\approx$~2 population.  Significantly improved
astrophysical insight into the AGN and host-galaxy properties of SMGs
will be provided with the next generation of observatories over the
coming decades. For example, mid-IR spectroscopy with the {\it James
Webb Space Telescope} ({\it JWST}; Gardner \etal 2006) will yield
constraints on the presence of broad emission lines in the rest-frame
near-IR band, potentially providing a direct route to estimating the
black-hole masses of the X-ray obscured SMGs, and the Atacama Large
Millimeter Array (ALMA; Kawakatu \etal 2007a; Wootten 2006) will
provide resolved dynamical measurements of the host galaxy down to
$\approx$~100~pc scales, including the circumnuclear region around the
black hole. If sufficiently sensitive optical and near-IR
spectropolarimeters can be developed on 30--50~m telescopes then it may
also be possible to directly search for the presence of broad emission
lines in scattered light in the X-ray obscured SMGs (e.g.,\ Antonucci
\& Miller 1985; Young \etal 1996). Finally, X-ray observations from
planned and proposed telescopes over the next $\approx$~10--15 years
will provide the potential to constrain the energetics of the AGN
activity, from either high signal-to-noise high-resolution spectroscopy
or photometric measurements out to X-ray energies of $\approx$~100~keV
[e.g.,\ the Japanese X-ray observatory {\it NEXT}; Ozawa \etal (2006),
the NASA observatories {\it Constellation-X} (White \etal 2004) and
{\it NuSTAR} (Harrison \etal 2005), and the ESA observatory {\it XEUS};
Parmar \etal (2006)].

\acknowledgements We gratefully acknowledge support from the Royal
Society (DMA; IRS), STFC (AMS; KC), NASA LTSA grant NAG5-13035 (WNB),
the {\it Chandra} Fellowship program (FEB), and the Alfred P. Sloan
Foundation and Research Corporation (AWB).  We thank M.~Volonteri for
insightful conversations, C.~Maraston for discussing stellar-mass
constraints based on her stellar evolution models, R.~McLure for
providing SDSS quasar data and useful feedback, and the anonymous
referee for a thoughtful report.

%%% Text of acknowledgements runs on after this command.

%%% THE BIBLIOGRAPHY
%%%
%%% CONSULT SECTION 3 OF "INSTRUCTIONS FOR AUTHORS" FOR HOW TO USE NATBIB.
%%% AUTHORS ARE ENCOURAGED TO USE EITHER THE "THEBIBLIOGRAPY" ENVIRONMENT
%%% BY UNCOMMENTING (DELETING THE "%" SYMBOL) THE COMMANDS BELOW, OR BY
%%% USING THE BIBTEX ENVIRONMENT. TO FIND OUT WHICH IS APPLICABLE TO YOUR
%%% CONTRIBUTION, CONSULT THE VOLUME EDITORS FOR YOUR PROCEEDINGS.
%%%

\newpage

\end{document}